\documentclass[aps,twocolumn,groupedaddress,showpacs]{revtex4}
\usepackage{epsf,latexsym}
\usepackage{times}

\newcommand{\bra}[1]{\langle #1 |}
\newcommand{\ket}[1]{| #1 \rangle}

\newcommand{\sandwich}[3]{\left \langle #1 | #2 | #3 \right\rangle}

\newcommand\N{{\mathrm {I\!N}}}

\newcommand{\be}{\begin{equation}}
\newcommand{\ee}{\end{equation}}
\newcommand{\bae}{\begin{eqnarray*}}
\newcommand{\eae}{\end{eqnarray*}}

\def\CC{{\rm\kern.24em \vrule width.04em height1.46ex depth-.07ex
    \kern-.30em C}}
\def\P{{\rm I\kern-.25em P}}

\def\bbbc{{\mathchoice {\setbox0=\hbox{$\displaystyle\rm C$}\hbox{\hbox
to0pt{\kern0.4\wd0\vrule height0.9\ht0\hss}\box0}}
{\setbox0=\hbox{$\textstyle\rm C$}\hbox{\hbox
to0pt{\kern0.4\wd0\vrule height0.9\ht0\hss}\box0}}
{\setbox0=\hbox{$\scriptstyle\rm C$}\hbox{\hbox
to0pt{\kern0.4\wd0\vrule height0.9\ht0\hss}\box0}}
{\setbox0=\hbox{$\scriptscriptstyle\rm C$}\hbox{\hbox
to0pt{\kern0.4\wd0\vrule height0.9\ht0\hss}\box0}}}}

\def\bbbz{{\mathchoice {\hbox{$\sf\textstyle Z\kern-0.4em Z$}}
{\hbox{$\sf\textstyle Z\kern-0.4em Z$}}
{\hbox{$\sf\scriptstyle Z\kern-0.3em Z$}}
{\hbox{$\sf\scriptscriptstyle Z\kern-0.2em Z$}}}}

\newcommand{\putfigs}[3]{$$\leavevmode\hbox{\epsfxsize=#2 cm \epsfysize=#3 cm 
\epsffile{#1.eps}}$$}

\newcommand{\putfig}[2]{$$\leavevmode\hbox{\epsfxsize=#2 cm
   \epsffile{#1.eps}}$$}

\begin{document}
\title{Mode Entanglement and Entangling power in Bosonic Graphs}
\author{Paolo Giorda$^{1,2}$ and Paolo Zanardi$^{1,2,3}$}
\affiliation{$^1$ Istituto Nazionale per la Fisica della Materia (INFM), UdR Torino-
Politecnico, 10129 Torino, Italy}
\affiliation{$^2$ Institute for Scientific Interchange (ISI), Villa Gualino, Viale 
Settimio Severo 65, I-10133 Torino, Italy}
\affiliation{$^3$ Department of Mechanical Engineering,
Massachusetts Institute of Technology, Cambridge Massachusetts 02139}

\begin{abstract}
We analyze the quantum entanglement properties of bosonic particles hopping over graph structures. 
Mode-entanglement of a graph vertex with respect the rest of the graph is  generated, starting from 
a product state, by turning on for a finite time a tunneling along the graph edges.
 The maximum achieved during the dynamical evolution 
by this bi-partite entanglement characterizes the entangling power of a given hopping hamiltonian.
We studied this entangling power  as a function of the self-interaction parameters i.e., non-linearities,
for all the graphs up to four vertices and for two different natural choices of the initial state.
The role of graph topology and self-interaction strengths in optimizing entanglement generation
is extensively studied by means of  exact numerical simulations and by perturbative calculations  
\end{abstract}

\maketitle

\section{introduction}
Generation of quantum entanglement is an essential prerequisite for the majority
of quantum information and computation tasks \cite{qip}.  To this aim many-body interactions
have to be  cleverly  engineered and carefully controlled in the variety of physical systems 
that pretend to represent viable candidates for quantum information processors.
The resulting quantum dynamics can be then characterized by means of its entangling capabilities
in different ways \cite{ep,ep_cirac,lea, man}. These {\em entangling power} measures provide a quantitive
to assess the dynamical strength \cite{man}  of a quantum evolution  as physical resource for quantum 
information processing (QIP) protocols. In this respect  the optimization of the entangling power
in the set of physically available interactions clearly represents a quite relevant  goal.

In this paper we will address this problem in the context of bosonic systems
on  lattice structures e.g., ultracold bosonic atoms in optical lattices.
This kind of systems have been already  rather extensively studied in the QIP related literature
\cite{milb_twowells, bec_lattice,chen, radu_pz,duan_ent, simon,hines,micheli,you}
in view of their experimental feasibility  and inherent great  flexibility 
that make them almost ideal candidates for studing and simulating \cite{duan_ctrl} a plethora of 
important   quantum many-body phenomena e.g., quantum phase transitions \cite{bec_mott}.

In the following  we will adopt the view of quantum entanglement
in systems of indistinguishable particles advocated in Refs \cite{virt,ind_ent,gitt, enk, shi, vedra}
in alternative to the complementary one pursued in Refs. \cite{schli,pask,Li}.
In the former the subsystems that get entangled are provided by bosonic {\em modes}
whereas in the latter are {\em particles} themselves. The mode-entanglement
approach is an intrinsecally  second-quantized one: the tensor product structure 
of the state-space necessary in order to define entanglement is provided
the natural identification of the bosonic (fermionic) Fock space with a set of linear oscillators
(qubits) associated
 with the single-particle wavefunctions.     

We shall analyze systems of bosonic particles hopping over inhomogneous spatial
structures \cite{bur,zan_graph} mathematically modelled by {\em graphs} \cite{graph}. The main objective
is to determine the role that the graph topology plays in the entangling power of the quantum dynamics
described by a Bose-Hubbard model \cite{bh}. 
For a given tunneling graph structure we will study
how the entangling capabilities are affected by the amount of non-linearity i.e., bosons self-interactions
present in the Hamiltonian. Numerical simulations as well as perturbative analytical arguments
will be provided for all the graphs up to four vertices. This analysis allows to optimzes the entanglement
production, from an initial product state, obtainable by switching on-and off the tunneling between
the various garph vertices i.e., one-body operator control, and by varying e.g., by Fesdbach resonances,
the strength   of the boson-boson local coupling. We believe that the theoretical framework
developed in this paper will have direct experimental relevance.

The paper is organized as follows: in Sect II is briefly recalled the Bose-Hubbard Hamiltonian
and in Sect III it will be applied to the case of hopping over spatially inhomogeneus structures
(graphs). In Sect IV we will describe the notion of bosonic mode-entanglement for the systems of
 interest and introduce the functional  for measuring the entangling power of a quantum  evolution.
Sect V is the major one, it contains the results of numerical simulations for all the graphs up
to four vertices. Finally Sect VI provides a summary and conclusions.

\section{The Bose-Hubbard model}

The systems we will study in this paper are all given by $L$ bosonic sites 
filled with $N=L$ bosonic particles. Some of the sites are linked to each other 
in the sense that the particles are allowed to tunnel from one site to another. 
The links existing between the sites will be described later (section III) by 
means of graphs. The particles in a site interact via a two-body interaction 
process (self-interaction).

The proper framework that can be used to describe the systems under study is 
given by the Bose-Hubbard (BH) model \cite{bh}. 
This model has been recently use to investigate many interesting properties of 
different systems going from arrays of BEC where the sites form an infinite 
(periodic) lattice \cite{bec_lattice}, to systems in which the number of sites 
is finite (dimer, trimer)\cite{milb_twowells}. 
Moreover this model has recently been used to describe new possible schemes for 
quantum computing \cite{radu_pz}.
According to this model the Hamiltonian of the system of $L$ sites and $N$ 
particles can be written as:
\be
H = \sum_i \varepsilon_i n_i^2 + \sum_i \gamma_i n_i  +\sum_{<i,j>} 
\tau_{ij}(c_i^\dagger c_j + h.c.) 
\ee
where: the indexes $i,j$ run in $V = \{0,1,2...,L-1\}$, the set of sites; 
$c_i^\dagger \mbox{ and } c_i$ are bosonic creation and annihilation operators 
that create/destroy a particle in the locale mode (site) $i$; 
$n_i = c_i^\dagger c_i $ are the corresponding occupation number operators.

The parameters $\gamma_i$ can be used to model the offsets in the ground state 
energies of the different sites. We suppose $\gamma_i = \gamma = 
\mbox{const} \  \forall i$ and from now on we neglect this term. In fact, 
since the total number of particles in our systems will be held fixed, the term 
$\gamma \sum_i n_i = \gamma N$ commutes with the Hamiltonian (1) and its 
contribution in the dynamics of the system consists only by a global 
phase.
The parameters $\varepsilon_i$ control the non-linear two-body interaction 
between the particles in each site $i$, while the parameters $\tau_{ij}$ control 
the tunneling processes between the sites $i,j$. We consider only 
tunneling between next-neighbourghring sites and we suppose to be able to 
set the tunnel parameters $\tau_{ij}$ and the self-interaction parameters 
$\varepsilon_i$ to some fixed desired values. These will be held fixed during 
the evolution, i.e. we will be intersted only on the free evolution of our 
systems. The latter will be described in detail in the next section.

\section{GRAPHS AND HAMILTONIANS}

As described in the previous section, each of the systems we consider is given by a lattice of $L$ vertices (sites or modes) filled with $N=L $
bosonic particles and we suppose that its Hamiltonian can be described in 
the framework of the 
Bose Hubbard model. We consider systems in which the self-interaction terms 
are all equal ($\varepsilon_i=\varepsilon \ \forall i$), while for the hopping 
parameters we consider the case $\tau_{ij} \in \{0,\tau \} \ \forall i,j $.  We focus 
on the free evolution of the simpliest of these systems, i.e. $N=L \in 
\{3,4\}$. 

For $L=N=3,4$ we consider only inequivalent systems, in the sense specified 
below,  each characterized by a different rooted graph $\Gamma=(V,E,r)$, where 
$V=\{0,1,2,..,L-1 \}$ is the set of vertices, $E= \{(i,j):i,j\in V,\ i \neq j \}$is the set of edges and $r \in V$ is the root of the graph. The latter will be considered as the mode (vertex) which is relevant for the calculation of the mode 
entanglement.
For each rooted graph $\Gamma$ the pair $(V,E)$ identifies a particular Bose 
Hubbard hamiltonian $H_{(V,E)}$, since the pair $(V,E)$ identifies a set of 
sites and the way in which these sites are connected. With this Hamiltionian one 
can build the evolution operator $U_{(V,E)}(t)=\exp(-itH_{(V,E)})$. 

If the pairs $(V,E)$ and $(V,E^*)$ are such that $E \neq E^*$ they give rise to 
two sets of inequivalent rooted graphs since the connectivity of the two graphs 
they identify is different.
Morover, for a given pair $(V,E)$ one can define a certain rooted graph by 
specifying the root site $r \in V$, but only some of the possible choices can be 
considered inequivalent. If two rooted graphs differ only for the 
root, they will be considered inequivalent if there is a difference in how the 
roots are directly connected to the other vertices of the graph or if there 
is a difference in the connectivity of the subgraph $\Gamma'=(V',E')$ 
constituted by the set of vertices $V'= V\backslash r$ and the set of edges $E'= 
\{(i,j):i,j\in V', i \neq j \}$. 

We now give an example  of the procedure used to identify the different 
inequivalent systems by 
starting with $N=L=3$.  In this case we will consider three different 
inequivalent rooted graphs (see Fig. 1a). 

\begin{figure}
\putfigs{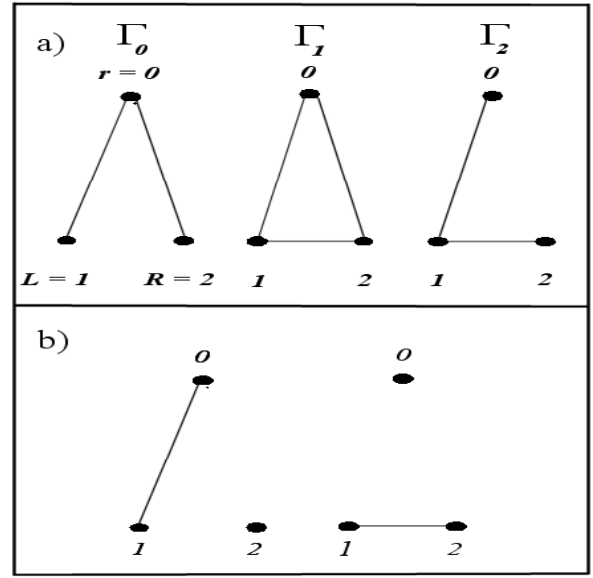}{8}{8}
\caption{$a)$ inequivalent rooted graphs for $N=L=3$; $b)$ disconnected graphs 
for $N=L=3$.}
\label{array}
\end{figure}

To build these graphs we first start by labeling the root vertex with 0, the left $(L)$ vertex with 1 and the right $(R)$ vertex with 2. The choice of the labels for the vertices of the subgraph $\Gamma'=(V', E')$, where $V'=\{L,R\}$ is irrelevant for our analysis (we could for example choose $L=2$ and $R=1$) since from the point of view of the entanglement of the root mode they are 
interchangeable and the relevant feature is how $r$ is directly connected to the 
vertices of the  subgraph $\Gamma'$ and the connectivity of the latter. 
So we can say that, for a given $E'$,the two labeling $(0,1,2)$ and  $(0,2,1)$ are equivalent and consider only 
the case $(r,L,R) \equiv(0,1,2)$. 

The rooted graphs $\Gamma_0$ and $\Gamma_2$ give rise to the same "kind" of BH 
Hamiltonians, in the sense that they both have two hopping terms; the only thing 
that changes 
is the labeling of the modes:
\[
\begin{array}{c}
H_{\Gamma_0}=\varepsilon(n_r^2+n_1^2+n_2^2)+\tau(c_r^\dagger c_1 + c_r^\dagger 
c_2 +\mbox(h.c.)) \\
 \downarrow (r \leftrightarrow 1) \\ 
H_{\Gamma_2}=\varepsilon(n_r^2+n_1^2+n_2^2)+\tau(c_r^\dagger c_1 + c_1^\dagger 
c_2 + \mbox(h.c.)).
\end{array}
\]
%$H_{\Gamma_1}=\varepsilon(n_r^2+n_1^2+n_2^2)+\tau(c_r^\dagger c_1 + c_1^\dagger 
%c_2 + \mbox(h.c.))+\tau( + \mbox(h.c.))
% \rightarrow (r \leftrightarrow 1) \rightarrow 
%H_{\Gamma_2}=\varepsilon(n_r^2+n_1^2+n_2^2)+\tau(c_r^\dagger c_1 + 
%\mbox(h.c.))+\tau(c_r^\dagger c_2 + \mbox(h.c.))$.
But, from the point of view of the root site, in $\Gamma_2 \  r$ is directly 
linked to one site (1), whereas 
in $\Gamma_0 \ r$ is directly linked with two sites (1,2); moreover the 
subgraphs have different connectivity: 
$E'_{\Gamma_2} \equiv \{(1,2)\} \neq E'_{\Gamma_0} \equiv \emptyset$.

The Hamiltonian based on $\Gamma_1$ differs from the other two since there are 
three hopping terms. In this case $r$ is directly connected to all the other 
sites as in $\Gamma_0$ but here the connectivity of the subgraph is different, 
$E'_{\Gamma_1}=\{(1,2)\}$.

Among all the possible graphs for the case $N=L=3$ we do not consider 
disconnected graphs like the ones in Fig.1b. The first case is a particular instance  of the system $L=2$. In the second case, since in our analysis the initial state will always be a product state belonging to the site occupation number basis, no entanglement can be genertated between the root mode and the others.

\begin{figure}
\putfigs{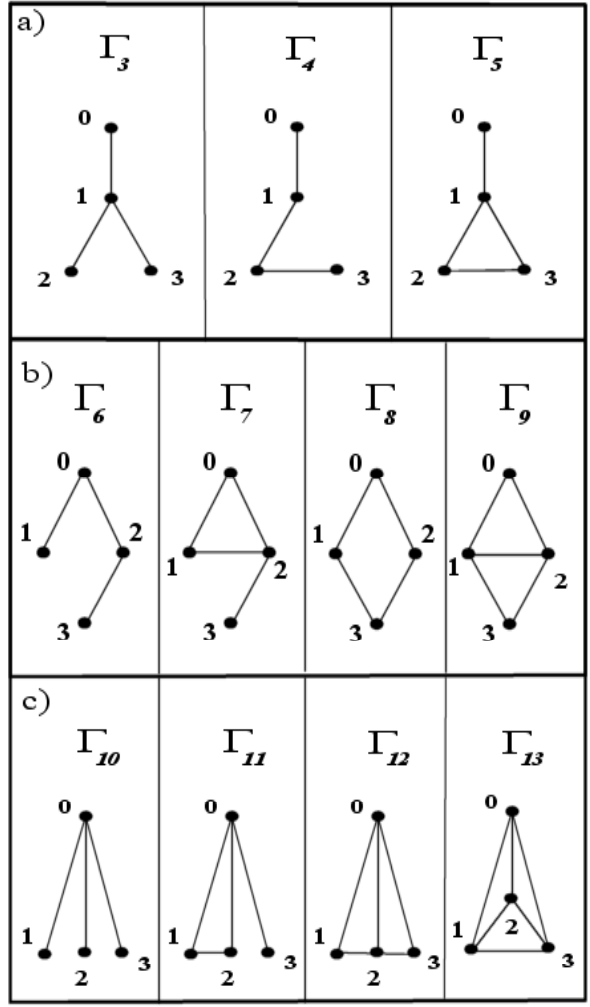}{8}{14}
\caption{Inequivalent rooted graphs for $N=L=4$; the root vertex $0$ is directly 
linked with 1 (a), 2 (b) and 3 (c) vertices.}
\label{array}
\end{figure}

Following the same ideas one can derive the inequivalent $\Gamma$'s also 
for the case $N=L=4$. One can start by choosing the root site and 
connect it to just one of the other vertices; then one has to find all the 
subgraphs $\Gamma'$ that give rise to inequivalent rooted graphs and repeat the 
whole procedure for the cases in which the root is directly connected with two 
and three of the other vertices. The result of this process is  given in Fig.2.
Once selected the relevant rooted graphs $\Gamma_j$ one can build the 
relative Hamiltonians $H_{\Gamma_j}(\varepsilon,\tau)$ and the relative 
evolution operators $U_{\Gamma_j}(t)=\exp(-itH_{\Gamma_j}(\varepsilon,\tau))$. 
The entagling power of the latter can then be calculated as described in the 
next section.

\section{MODE ENTAGLEMENT}

The problem of the entanglement of two-mode (component) Bose Einstein 
condensates (BEC) has been recently studied 
in different ways \cite{milb_twowells, micheli, duan_ent,hines}. In Ref. \cite{micheli} the system is composed by a 
two component BEC consisting  of N atoms that can access two different hyperfine states $\ket{A}$ and 
$\ket{B}$ coupled by an electromagnetic field. In this paper
it is stated that the maximally entangled state for the system is 
given by 
\[\ket{GHZ}_N=\frac{1}{\sqrt{2}}(\ket{A}^{\otimes^N}+\ket{B}^{\otimes^N})\] 
which is 
the N-partite generalization of the GHZ (Greenberg- Horne-Zeilinger) state 
for 3 qubits 
\[\ket{GHZ}_N=\frac{1}{\sqrt{2}}(\ket{000}+\ket{111}).\]  
The underlying decomposition of the system is defined by considering  each of 
the N bosons as a subsystem 
living in the two dimensional Hilbert space given by 
$\mbox{span}\{\ket{A},\ket{B}\}$.

In this picture to each boson   corresponds 
a qubit and the Hilbert space of the total system then corresponds to the 
Hilbert space of N qubits.
On the other hand, as it has been pointed out in \cite{hines}, the idea to consider 
different bosons  as different two-levels subsystems and to calculate the value of the 
entanglement according to some suitable measure, althought mathematically possible, it is physically questionable in 
view of the indistinguishability  of the particles. 
In order to have some entanglement between two systems, these have to be separately accessible from the experimental point of view. 

In general one can think to 
speak of the entanglement between  indistinguishable particles, but it is not possible to make any use of this 
resource since there are no  physically realizable measurements on the system that can discriminate between 
the particles. 

In \cite{milb_twowells}   one of the systems under study is a pair of tunnel-coupled BEC's 
considered as a realization of a bimodal BEC. In this paper, On the contrary of what is done in Ref \cite{micheli}, in order to 
calculate the entanglement present in the ground state of the system, the subsystems that are considered 
physically meaningful are the two modes. The number of  bosons present in the system is 
always fixed. 
In this set up it is possible to have physical access to the subsystems. For 
example, one realization of the model  is given by a BEC with a fixed number of bosonic 
particles that are trapped in a two wells potential, where each well (site) can be considered as a 
mode; here the modes are spatially separated. The indistinguishability of the particles does not allow to 
determine experimentally in which mode is a specific particle, but it is physically possible to 
measure the occupation number of each mode, i.e. the number of particles present at a given instant in 
the wells. The system can be viewed as a bipartite system for which a good measure of the entanglement exist: 
the Von Neumann entropy of the reduced density matrix. In this paper we follow this point of view 
and we make use of the notion of mode entanglement to characterize our systems. 

We consider systems in which the 
number of modes (sites) $L$ is fixed and is always equal to the number of 
particles $N\in\{3,4\}$. 
The Hilbert space $H^{(N)}$ of the system for a fixed $N=L$ is a subspace of  
the Fock space $h^{\otimes L}$, where 
$h=\mbox{span}\{\ket{n }\}_{ n \in \N}$.
% and
%\[
%  \Lambda = \{\vec{n}=(n_0,n_1,..,n_{L-1})\}_{n_0,n_1,..,n_{L-1}=0}^\infty.
%\]
The subspace can be written as 
\[
H^{(N)}=\mbox{span}\{\otimes_{j=0}^{L-1} \ket{n_j}=\ket{\vec{n}}: \vec{n} \in 
\Lambda^N \}
\]
where
\[
 \Lambda^N=\{\vec{n}=(n_0,n_1,..,n_{L-1}): n_j \in \N \mbox{ and } 
\sum_{j=0}^{L-1} n_j=N \}.
\] 
The general state of the system can then be written as $\ket{\psi}= 
\sum_{\vec{n} \in \Lambda^N }c(\vec{n})\ket{\vec{n}}$.

We are interested in calculating the entanglement 
of one of the modes, say the mode 0, with respect to the rest of the system. In 
practice we can always 
reduce to a bipartite picture and we make use of the Von Neumann entropy to 
calculate the entanglement 
of the relevant mode. To do this we need to calculate the on-site reduced 
density matrix. It is interesting 
to note that $H^{(N)}$ cannot be written in terms of the tensor product of the 
Hilbert space of the mode subsystems: 
in order to calculate the on-site reduced density matrix one has to remind that 
the $H^{(N)}$ is a subspace of  
the entire Fock space, that by definition is written as a tensor product of the 
mode subsystems. Thus  one can borrow 
this tensor product structure in order to calculate the trace over all the modes 
but the 0 one.  Let be $V=\{0,1,..,L-1\} $ the set of sites (modes, or vertices 
in the graphs terminology) 
then the reduced density matrix of the site 0 can be written as:
\begin{eqnarray*}
\rho^{(0)} & = & \mbox{Tr}_{V_{\backslash 0}}(\ket{\psi}\bra{\psi}) \\
& = & \sum_{\vec{n}\in \Lambda_{\backslash 0}}
\bra{\vec{n}} \left( \sum_{\vec{h},\vec{k} \in \Lambda^N}             		
c(\vec{h})\, c^*(\vec{k})\ket{\vec{h}}\bra{\vec{k}} \right)\ket{\vec{n}}  	\\
& = & \sum_{j=0}^N \ket{j}\bra{j} \left( \sum_{\vec{n}(j)} |c(\vec{n}(j))|^2 \right) \\
& = & \sum_{j=0}^N
\rho^{(0)}_j\ket{j}\bra{j}
\end{eqnarray*}
where 

\[
\Lambda_{\backslash 0} = \{\vec{n}=(n_1,..,n_{L-1}) \ : \ n_j \in \N \},
\]
and where $\vec n(j)=(j,n_1,..,n_{L-1})\in \Lambda^N$.

In view of the constrain given by the fixed number of particles $\rho^{(0)}$ is 
diagonal and the Von Neumann entropy 
relative to the mode 0 can be simply written as: 
\begin{eqnarray*}
e^{(0)} & =& S(\rho^{(0)})=-Tr(\rho^{(0)} \log_2\rho^{(0)}) \\
& = & -\sum_{j=0}^N \rho^{(0)}_j \log_2\rho^{(0)}_j.
\end{eqnarray*}
That is, $ e^{(0)}$ is the Shannon entropy of the probability distribution
\be
\{ \rho^{(0)}_0, \rho^{(0)}_1,...., \rho^{(0)}_N \}.
\ee

In order  to characterize the different systems under consideration, and that 
are described in detail 
in the next section , we use of the following functional
\[
EP^{(0)} = [\sup_{t\in[0,T]}(e^{(0)}(t))]/\log(N+1)
\]
The expression $\sup_{t\in[0,T]}(e^{(0)}(t))$ represents the maximum value 
of the entanglement of site 0 with respect to the rest of the system during the evolution of the system; this value has been nomalized with respect to 
$\log(N+1)$ that is the maximum value of the mode entanglement that can be 
reached by the system and that corresponds to the situation in which 
$\rho^{(0)}=(\frac{1}{N+1})I$. 
We take the value of $EP^{(0)}$ as a measure of the entagling power 
which  characterizes the pair $(U_{\Gamma_j},\ket{\psi_{in}})$; what we want to 
compute is, for each $\Gamma_j$, the entangling 
capability of the evolution operator $ U_{\Gamma_j}$ when it is applied for a given period of time to the initial state $\ket{\psi_{in}}$. 

As far as the choice of the initial states is concerned, we decide to use 
two particular kind of states for both the case $N=L=3$ and $N=L=4$. The first 
kind of states are the uniformly occupied ones : $\ket{111},\ket{1111}$. The 
second kind of states are the ones in which all the particles are localized in 
the mode $0$:
$\ket{300},\ket{4000}$. This states allow us to test the entangling power of the 
operators $ U_{\Gamma_j}$ in two opposite and in a sense "extremal" situations 
with respect to the localization of the particles in the sites.

This choice is also motivated by the possible physical implementations of the 
model under study. In fact one of these possible physical realizations is 
given by a set of ultracold bosons placed in an optical potential.
The goal of placing exactly one atom in each well of an optical lattice created 
by a set of counterpropagating laser beams has been recently achieved 
\cite{bec_mott} thanks to the realization of a super-fluid/Mott-insulator phase 
transition. This has motivated us to choose, among all the possible initial 
states of the evolution, the ones  for which the site occupation number is equal 
to one, $(n_j=1,\ \forall j=0,..,L-1)$. In this kind of application, in order 
to realize the actual systems we are interested in, one should be able to 
isolate the relevant sites with the required graph topology. 

The choice of the states in which all the particles are localized in one single 
mode is motivated by the fact that the typical realization of a BEC consists in 
the localization of all the bosons in a trapping (parabolic) potential; with a 
further step it is possible to create a lattice, by means of properly sized 
counterpropagating laser beams. In this kind of realization one should 
be able to isolate the relevant sites with the required graph topology and to 
fill with the proper number of particles the desired mode (well, site). 

In general, for the realization of our model, one should also be able to 
manipulate the control parameters of the model, that is to fix the 
self-interaction and hopping parameters to the desired values, that are the same 
for all the sites and all the links between them. Thus, althought the organization 
of the sites in graphs could be a difficult task, the realization of our model 
requires only a "global" control of the parameters since it does not require to 
perform  control operations on single sites and/or single links between 
sites.

\section{RESULTS}
Our goal is to characterize, for each $(U_{\Gamma_j},\ket{\psi_{in}})$, the behaviour of the entangling power $EP^{(0)}_{\Gamma_j}(\varepsilon,\tau)$ when the ratio between the self-interaction parameter  and the hopping parameter $\varepsilon / \tau$ varies from zero to a value that is much greater then one. In order to do so we fix $\tau = 1$ (that is we measure $\varepsilon$ in  $\tau$ units) and let $\varepsilon$ vary in $[-\varepsilon_{max},\varepsilon_{max}]$ ($\varepsilon_{max} > 0$) with step $\triangle \varepsilon$.
For each $\varepsilon \in [-\varepsilon_{max},\varepsilon_{max}]$ we simulate the full quantum evolution for $t \in [0, T]$ that the system undegoes according to the unitary operator $U_{\Gamma_j}(t)=\exp(-tH_{\Gamma_j}(\varepsilon,\tau))$, $(\hbar=1)$, when the initial state is $\ket{\psi_{in}}$.
The evolution time $T$ has to be taken much greater than all the 
eigen-periods of the syetems i.e., $T\gg \mbox{max}_i E^{-1}_i,$ where
$E_i$ denote an energy eigenvalue. If this latter condition holds one has 
no changes in the entangling power by allowing a longer  evolution time.
We explicitly verified this asymptotic behaviour in our numerical 
simulations.
 For each of these evolutions we calculate the value of $EP^{(0)}_{\Gamma_j}$ .

In this way we can consider three different regimes. When $\varepsilon = 0$ there is no self-interaction between the bosonic particles but these are free to hop between the different sites according to the graph $\Gamma$. In this regime the intial state $\ket{\psi_{in}}$ is not an eigenstate of the pure hopping Hamiltonian. 
When $\varepsilon \rightarrow \varepsilon_{max}$ we have $\varepsilon / 
\tau \gg 1$, the hopping between the different sites is highly reduced and the 
systems tend to evolve toward states which are very close to the initial ones 
$\ket{\psi_{in}}=\ket{\vec{n}}$, that in this case are eigenstates of the pure 
self-interaction Hamiltonians corresponding to the eigenvalues $\varepsilon 
\sum_{i=0}^{L-1} n_i^2$.
The last regime is the intermediate one in which there is a competition of the 
two processes that govern the evolution, self-interaction and hopping.

To simulate the evolution of the systems under study we use a second order 
Runge-Kutta routine to integrate the time dependent Shrodinger equation for the 
different Hamiltonians.

\begin{figure}
\putfig{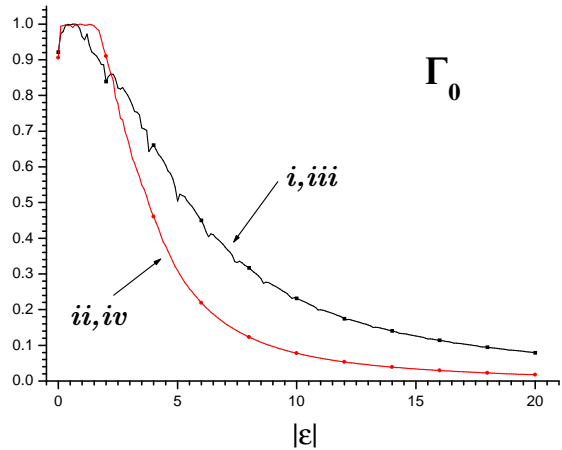}{10}
\caption{Plot of $EP_{\Gamma_{0}}^0 (|\varepsilon|)$ for 
$\ket{\psi_{in}}=\ket{111} \ (i \mbox{ and } iii)$ and for 
$\ket{\psi_{in}}=\ket{300}\ (ii \mbox{ and } iv) $; the curves $i \mbox{ and } 
ii$ refer to $|\varepsilon|=\varepsilon$, while the curves $iii \mbox{ and } vi$ 
refer to $|\varepsilon|=-\varepsilon$.}
\label{array}
\end{figure}

\begin{figure}
\putfig{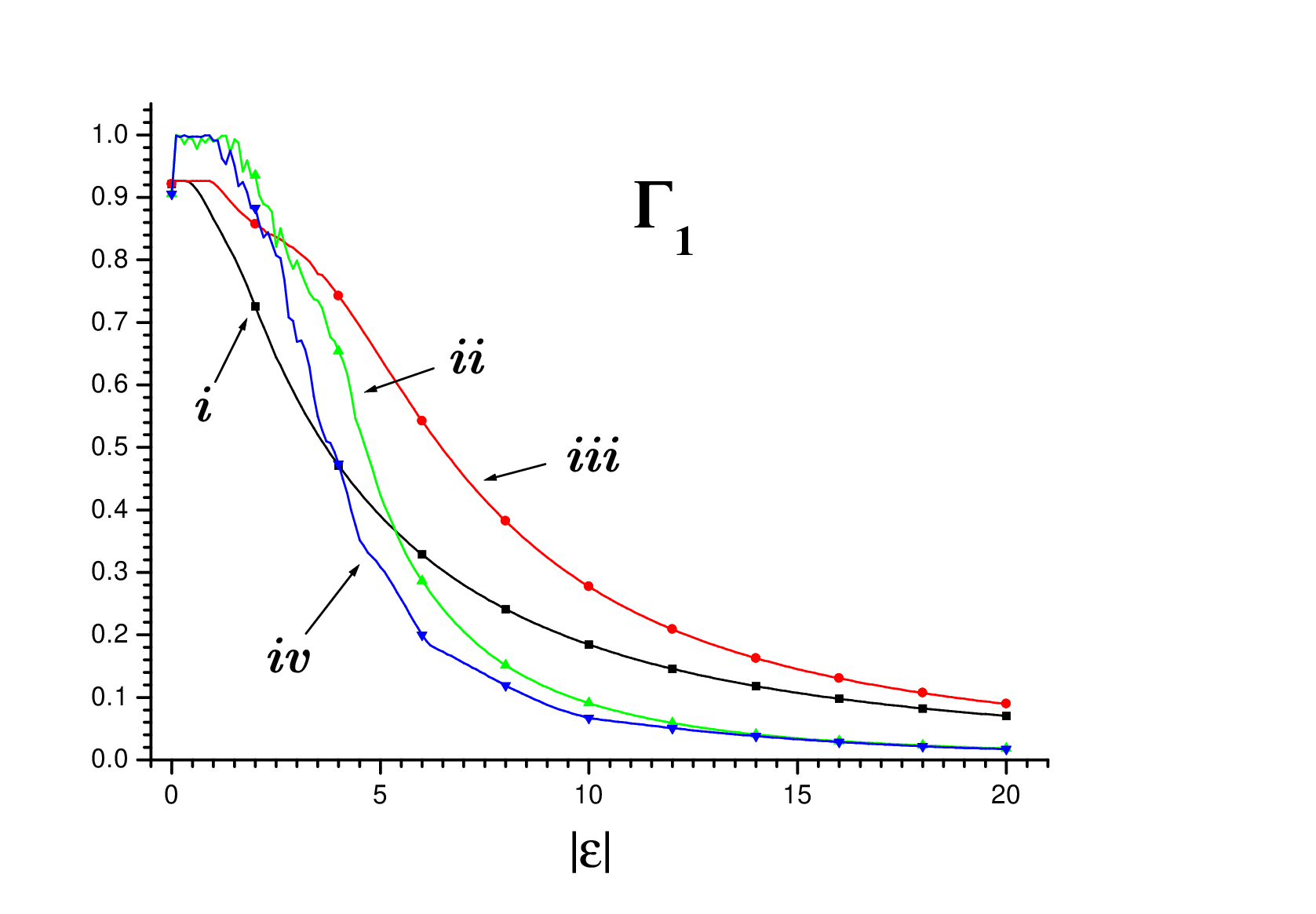}{10}
\caption{ Plot of $EP_{\Gamma_{1}}^0 (|\varepsilon|)$ for 
$\ket{\psi_{in}}=\ket{111} \ (i \mbox{ and } iii)$ and for 
$\ket{\psi_{in}}=\ket{300}\ (ii \mbox{ and } iv) $; the curves $i \mbox{ and } 
ii$ refer to $|\varepsilon|=\varepsilon$, while the curves $iii \mbox{ and } vi$ 
refer to $|\varepsilon|=-\varepsilon$.}
\label{array}
\end{figure}

\begin{figure}
\putfig{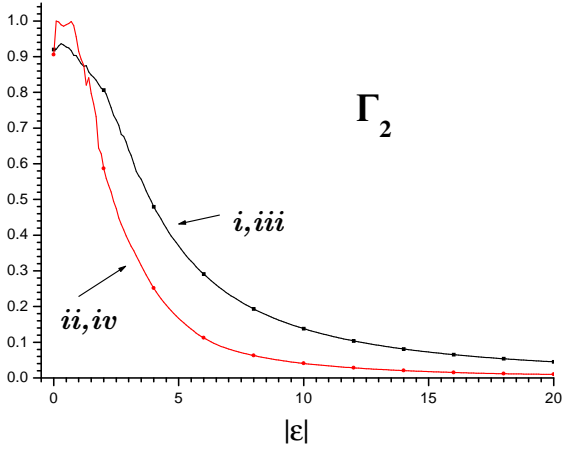}{10}
\caption{ Plot of $EP_{\Gamma_{2}}^0 (|\varepsilon|)$ for 
$\ket{\psi_{in}}=\ket{111} \ (i \mbox{ and } iii)$ and for 
$\ket{\psi_{in}}=\ket{300}\ (ii \mbox{ and } iv) $; the curves $i \mbox{ and } 
ii$ refer to $|\varepsilon|=\varepsilon$, while the curves $iii \mbox{ and } vi$ 
refer to $|\varepsilon|=-\varepsilon$.}
\label{array}
\end{figure}

\subsection{N=L=3}
In this subsection we illustrate the results of our simulations for the case 
$N=L=3$. In the following we refer to the figures 3, 4 and 5. For each graph 
$\Gamma_j, j\in\{0,1,2\}$ we have computed $EP^{(0)}_{\Gamma_j}$ as a function of $\varepsilon $, using  $\triangle \varepsilon = 0.1$. For each value of 
$\varepsilon$ we evolved the system from $t=0$ to $T=15$ (arbitrary 
units).
For the curves $(i)$ and the $(iii)$ the intial state is $\ket{\psi_{in}}=\ket{111}$. The curves
$(ii)$ and the $(iv)$ refer to simulations in which the initial state is 
$\ket{\psi_{in}}=\ket{300}$. In the $(i)$ and $(ii)$ cases $\varepsilon$ varies in the interval $[0,\varepsilon_{max}]$, while in the 
$(iii)$ and  $(iv)$ cases $\varepsilon\in [-\varepsilon_{max},0]$; for all the 
different cases $\varepsilon_{max}=+20$ and the value of $EP^{(0)}_{\Gamma_j}$ has been plotted in function of $|\varepsilon|$.

\vspace{0.2cm}

A first interesting feature about the systems under study is that when 
$\varepsilon = 0$ the entagling power of the evolution operator $U_{\Gamma_j}$ 
does not depend on the graph $\Gamma_j$ but only on the initial state. 
In fact, $EP^{(0)}_{\Gamma_j} \cong 0.92,\  \forall j $, in the case 
$\ket{\psi_{in}}=\ket{111}$ while $EP^{(0)}_{\Gamma_j} \cong 0.91, \ \forall j$, 
for $\ket{\psi_{in}}=\ket{300}$. The entagling capability of the 
$U_{\Gamma_j}$'s , when $\varepsilon = 0$, is then only slightly enhanced if the 
system starts in a situation in which the particles are uniformly distributed 
over the sites.

\vspace{0.2cm}

We now turn to analize the maximum values reached by $EP^{(0)}_{\Gamma_j}$ for 
the different sistems. A first fact that emerges from our simulations is 
that the maximum theoretical value of the mode entanglement $(\log_2 4)$ can be 
reached for any $j$  if the system starts from $\ket{\psi_{in}}=\ket{300}$. In fact, for an appropriate choice of the ratio $|\varepsilon| / \tau$, regardless of the sign of the self-interaction, $EP^{(0)}_{\Gamma_j}(\ket{300},\varepsilon) > 0.99$. If $\ket{\psi_{in}}=\ket{111}$ this is true 
only in the case $j=0$. Another interesting feature, common to the cases just 
mentioned, is that $EP^{(0)}_{\Gamma_j}$ increases very rapidly when 
$|\varepsilon|$ starts to differ from $0$ and passes from $0.91-0.92$ to values 
$> 0.97$ in correspondance of $|\varepsilon| = 0.1$; then the value of 
$EP^{(0)}_{\Gamma_j}$ remains very high $(>0.95)$ for $|\varepsilon| < 
\varepsilon'$, where $\varepsilon'>0$ differs for the various cases. We can the 
say that, in this interval of values and for the above mentioned choice of 
$(\Gamma_j, \ket{\psi_{in}})$, the self-interaction  between the particles, 
either repulsive or attractive, enhances the capability of the $U_{\Gamma_j}$'s 
to entangle the mode $0$ with the rest of the system. 

On the contrary, in the cases $(\Gamma_1, \ket{111})$ and $(\Gamma_2, \ket{111})$ the maximum theoretical value of the mode entanglement is never reached and turning on the self-interaction only slightly enhance the entangling capability of the $U_{\Gamma_j}$'s; this happens in correspondance of values of $|\varepsilon|$ which are smaller or of the same order of $\tau (=1)$.

\vspace{0.2cm}

Another interesting feature highlighted by our simulations is related to the 
change of sign of the self-interaction parameter $(\varepsilon \rightarrow -
\varepsilon)$ and a particular property of the graphs $\Gamma_j$'s.
If we compare the simulations $i$'s vs $iii$'s we can see that for the graphs 
$\Gamma_0$ and $\Gamma_2$ 
\[
EP^{(0)}_{\Gamma_j}(\varepsilon)=EP^{(0)}_{\Gamma_j}(-\varepsilon)\,\   \forall 
\varepsilon \in [0,20] 
\]
while for $\Gamma_1$ the curves $i$ and $iii$ differ significantly; the same 
holds for the curves  $ii$ and $vi$. The difference of the behaviour of 
$EP^{(0)}_{\Gamma_j}$ in the two cases depends on a particular property of the graphs we have 
used to characterize our systems. These can be distinguished in 
{\it bi-partite} and  {\it non bi-partite} graphs. A graph $\Gamma$ is bi-partite if
\[
V= A \cup B \mbox{ and  } (a,b) \in E \Leftrightarrow a \in A \mbox{ and } b \in 
B.
\]
Suppose for example to colour each of the vertices of a graph with one of two 
different colours $A$ and $B$. A graph is bi-partite if the vertices coloured 
with $A$ are linked only with vertices coloured with $B$, and viceversa. In our 
case the bi-partite graphs are $\Gamma_0$ and $\Gamma_2$ (note 
that the bi-partiteness of a graph depends only on the pair $(V,E)$ and is 
independent from the root).

If $\Gamma_j$ is bi-partite, the Hamiltonian that results from 
the change of sign of $\varepsilon$ can be written as 
\be
H_{\Gamma_j}(-\varepsilon , \tau)=-P \cdot H_{\Gamma_j}(\varepsilon , \tau)\cdot 
P^\dagger,
\ee
where  $P=\exp(-i \pi \sum_{j \in A} n_j)$ is the unitary operator that 
implements the canonical transformation 
\be
c_j \rightarrow (-1)^{\chi_A(j)} c_j
\ee
($\chi_A$ is the characteristic function on the subgraph $A$). The evolution 
operator  for a given bi-partite graph can then be expressed as
\bae
U_{\Gamma_j}^{-\varepsilon}(t)& = &\exp(-itH_{\Gamma_j}(-\varepsilon , \tau)) =  
P \exp(it H_{\Gamma_j}(\varepsilon , \tau)) P^\dagger \\ & = & 
P U_{\Gamma_j}^{\varepsilon}(-t) P^\dagger
\eae

If we now apply this operator to a state belonging to the occupation number 
basis $\ket{\vec n}$ we obtain:
\bae
\ket{\psi_{\vec n}^{-\varepsilon} (t)} & = & U_{\Gamma_j}^{-\varepsilon}(t) 
\ket{\vec n} 
= P \exp(it H_{\Gamma_j}(\varepsilon , \tau))\exp(-i\phi_{\vec n})\ket{\vec n} 
\\
& = & \sum_{\vec m \in \Lambda^N}\exp(i\phi_{\vec m}) \alpha_{\vec n \vec m}(-t) 
\exp(-i\phi_{\vec n})\ket{\vec m}
\eae

where $\exp(-i\phi_{\vec n})$ and $\exp(i\phi_{\vec m})$ are, respectively, the 
phase factors given by the action of the operator $P^\dagger$ on the initial 
state $\ket{\vec n}$ and of the operator $P$ on the states $\ket{\vec m}$ in the 
sum.
The terms $\alpha_{\vec n \vec m}(-t)$ can be obtained by  acting with the 
time reversed operator $U_{\Gamma_j}^{\varepsilon}(-t)$ on $P^\dagger \ket{\vec 
n}$ when it is written in terms of the energy basis states, in which the 
Hamiltonian is diagonal, and then going back to the site occupation number 
basis. 
Since $\forall j$ $H_{\Gamma_j}$ is a real simmetric matrix there exist a real 
orthogonal matrix $O$ ($O \cdot O^T= 1$) that diagonalize it and with which it 
is possible to implement the required changes of basis. We can then write
\[
 \alpha_{\vec n \vec m}(-t) = \sum_k \beta_{\vec n \vec m k} \exp(+itE_k ), \ 
\forall \vec n , \vec m 
\]
where $E_k$ are the eigenvalues of the Hamiltonian and $\beta_{\vec n \vec m k}$ 
can be written in terms of the real entries of $O$. This fact is crucial when we 
compare the evolutions of $\ket{\vec n}$ obtained by applying $U_{\Gamma_j}^{-
\varepsilon}(t)$ and $U_{\Gamma_j}^{\varepsilon}(t)$ (note the sign of $t$). The 
latter gives:
\bae
\ket{\psi_{\vec n}^{\varepsilon} (t)} & = & \exp(-itH_{\Gamma_j}(\varepsilon , 
\tau)) \ket{\vec n} \\
& = & \sum_{\vec m \in \Lambda^N} \alpha_{\vec n \vec m}(t) \ket{\vec m}
\eae
where
\[
\alpha_{\vec n \vec m}(t)= \sum_k \beta_{\vec n \vec m k} \exp(-itE_k ), \ 
\forall \vec n , \vec m .
\]
We have now the elements to explain why,
when $\Gamma_j$ is bipartite, the curves $(i)$ and $(ii)$ overlap perfectly with 
the curves $iii)$ and $vi)$ respectively. Since  the terms $\beta_{\vec n \vec m 
k}$ are real we have that for any $t$ and $\forall \vec n , \vec m$ the 
amplitudes of the states $\ket{\vec m}$ in the expression of $\ket{\psi_{\vec 
n}^{-\varepsilon} (t)}$ and $\ket{\psi_{\vec n}^{\varepsilon} (t)}$ satisfy:
\be
|\exp(i\phi_{\vec m}) \alpha_{\vec n \vec m}(-t) \exp(-i\phi_{\vec 
n})|^2=|\alpha_{\vec n \vec m}(t)|^2,
\ee
in fact $\alpha_{\vec n \vec m}(-t)=\alpha_{\vec n \vec m}^*(t)$.
Consequently, since the value of the mode entanglement and of 
$EP^{(0)}_{\Gamma_j}$ depends only on the modulus square of the probability 
amplitudes of the basis states $\ket{\vec m}$ in the expression of 
$\ket{\psi_{\vec n}^{\pm \varepsilon} (t)}$ (see section IV), for any 
given $\ket{\psi_{in}}=\ket{\vec n}$ and  for all the bi-partite graphs 
we have $EP^{(0)}_{\Gamma_j}(\varepsilon)=EP^{(0)}_{\Gamma_j}(-\varepsilon)$. In 
general, this is no longer true for the non-bipartite graphs.

\vspace{0.2cm}

Another relevant and general feature of the entagling power $\forall \Gamma$'s 
and for both the selected initial state is that $EP^{(0)}_{\Gamma_j}$ rapidly 
decreses when $|\varepsilon|$ is significantly greater than $1 (=\tau)$ with a 
rate that is higer in the case $\ket{\psi_{in}}=\ket{300}$. This can be 
explained with the following argument.
When $|\varepsilon|$ becomes significantly greater then 1, the Hamiltonian of the system, $\forall \ \Gamma_j$, can be considered as composed of an unperturbed diagonal term $H_\varepsilon$, the self-interaction one, and a perturbation term given by the hopping matrix $H_\tau$. Both the initial states of our computation are eigenvectors of the unperturbed term and correspond to the eigenvalues $3\varepsilon$ ($\ket{111}$) and $9\varepsilon$ ($\ket{300}$). As $|\varepsilon|$ increses, the perturbation becomes smaller and the only processes relevant for the evolution are the first order ones. This means that 
the probability distribution  
$(\rho^{(0)}_0,\rho^{(0)}_1,\rho^{(0)}_2,\rho^{(0)}_3)$, given by the diagonal 
elements of $\rho^{0}$, is higly unbalanced and this results in a value of 
$e^{(0)}(t)$ that for any $t$ is very small, and so it is, consequently, the 
value of $EP^{(0)}_{\Gamma_j}$. 
 
To give a more detailed explanation of this phenomenon let's take for example 
$\Gamma_0$ and suppose $\ket{\psi_{in}}=\ket{111}$. In general, at first order, 
the state $\ket{\psi_{in}}$ can be coupled  by the perturbation to the states 
$\ket{\vec n'} \in \chi'_{ \ket{ \psi_{in} }}$, where
\[
\chi'_{ \ket{ \psi_{in}}} = \{\ket{ \vec n'} \ \ |\ \  \vec n' \in \Lambda^N 
\mbox{and}  \sandwich{\vec n'}{H_\tau}{\psi_{in}} \neq 0 \}.
\]
In our case $\chi'_{ \ket{111}}=\{\ket{021},\ket{012},\ket{201},  \ket{210}  
\}$:
the modulus square of the probability amplitudes of these states 
contribute to the terms $\rho^{(0)}_0,\rho^{(0)}_2 $ of the distribution. The state in 
which $n_0=3$ is coupled to $\ket{111}$ only by second order processes and for 
this reason the term $\rho^{(0)}_3 \ll \rho^{(0)}_0,\rho^{(0)}_1,\rho^{(0)}_2$; 
furthermore, when the perturbation is very small, the terms 
$\rho^{(0)}_0,\rho^{(0)}_2 \ll \rho^{(0)}_1$ and the distribution is very peaked 
around $\rho^{(0)}_1$. 

The same arguments hold for $\Gamma_0$ when $\ket{\psi_{in}}=\ket{300}$, 
but there is a significant difference between the two cases. In fact  $\chi'_{ 
\ket{300}}=\{\ket{201}, \ket{210} \}$ so that when the pertubation is small 
$\rho^{(0)}_0,\rho^{(0)}_1 \ll \rho^{(0)}_2 \ll \rho^{(0)}_3$: the probability 
distribution  is more unbalanced than in the previous case and this gives a 
first explanation of the fact that in the perturbative regime 
\[
EP^{(0)}_{\Gamma_j}(\ket{111}) > EP^{(0)}_{\Gamma_j}(\ket{300}), \ \forall j.
\] 

There is another important difference between the two cases that justifies this 
relation: the strenght of the coupling. This, at first order, depends on the 
ratio $\tau / \triangle E $, where $\triangle E $ is the  difference in energy 
between $\ket{\psi_{in}}$ and the states belonging to  $\chi'_{ \ket{ 
\psi_{in}}}$.  The ratio is higher for $\ket{\psi_{in}}=\ket{111} \ \ 
(1/2|\varepsilon|)$ then  for  $\ket{\psi_{in}}=\ket{300} \ \  
(1/4|\varepsilon|)$ and this results in higer  values of probability relative 
to the states in $\chi'_{ \ket{ \psi_{in}}}$ in the first case. Consequently the 
associated eigenvalues of $\rho^{(0)}$ can be higher for 
$\ket{\psi_{in}}=\ket{111}$ thus giving rise to a probability distribution which 
is flatter and to higher values of $EP^{(0)}_{\Gamma_j}$.

\vspace{0.2cm}

We now use the perturbative approach to describe the features relative to the 
non-bipartite graph $\Gamma_1$. Here the interesting thing to notice is again 
the role of the sign of the self-interaction, which depends on the initial state. 
For $\ket{\psi_{in}}=\ket{300}$ a repulsive (positive) self-interaction enhances the entangling capability of $U_{\Gamma_j}(t)$. In fact, for a large part of the 
interval $ [0,|\varepsilon_{max}|]$, 
\be
EP^{(0)}_{\Gamma_1}(\ket{300},+|\varepsilon|) \geq 
EP^{(0)}_{\Gamma_1}(\ket{300},-|\varepsilon|).
\ee
On the other hand, when
$\ket{\psi_{in}}=\ket{111}$ is the attractive (negative) self-interaction that 
favors the entangling power of the evolution operator. In fact for any 
$|\varepsilon| \in [0.5,\varepsilon_{max}]$
\be
EP^{(0)}_{\Gamma_1}(\ket{111},-
|\varepsilon|)>EP^{(0)}_{\Gamma_1}(\ket{111},+|\varepsilon|).
\ee
This behaviour can be understood in detail at least in a perturbative picture. 
Let's start from the case $\ket{\psi_{in}}=\ket{111}$. When $|\varepsilon|$ is 
significanly greater than $1 (=\tau)$ we can use a two-level picture to describe 
our system. The first level is given by  $\ket{\psi_{in}}$, while the second one 
is given a uniform superposition of the  $q = |\chi'_{ \ket{ \psi_{in} } }|$ 
degenerate states that are coupled to  $\ket{\psi_{in}}$ by first order 
processes, that is 
\be
\ket{I} = (1/\sqrt q ) \sum_{\ket{\vec n'} \in \chi'_{ \ket{ \psi_{in} } }} 
\ket{\vec n'} \propto H_\tau \ket{\psi_{in}}.
\ee
The two-levels Hamiltonian  can be written in general as 
\be
H= \left( \begin{array}{ccc}
E_1 & W \\
W & E_2
\end{array} \right)
\ee
where
\bae
E_1 & =& \sandwich{\psi_{in}}{H_\varepsilon}{\psi_{in}}, \\
W & = & \sandwich{I}{H_\tau}{\psi_{in}} = \sandwich{\psi_{in}}{H_\tau}{I}^*, \\
E_2 & = & \sandwich{I}{H_\varepsilon}{I} + \sandwich{I}{H_\tau}{I}.
\eae
It is a standard result of two coupled levels systems theory that the 
probability of finding the system in the state $\ket{I}$  at time $t$ can be 
written as (Rabi's formula)
\bae
P_{\ket {\psi_{in}} \rightarrow \ket{I}}(t) & = & |\sandwich{I}{\exp[-
it(H_\varepsilon + H_\tau)}{\psi_{in}}|^2 \\
 & = & \frac{4|W|^2}{4|W|^2 + (E_1 - E_2)^2} \sin^2 (\theta t), 
\eae
where $\theta = \sqrt {4|W|^2 + (E_1 - E_2)^2}/2$.
Here the important fact is the role of the term $(E_1 - E_2)^2$ in the 
denominator from which depends the maximum  value attainable by $P_{\ket 
{\psi_{in}} \rightarrow \ket{I}}(t)$; this is directly linked with the maximum values of  
$\rho^{(0)}_0,\rho^{(0)}_2 $ and consequently with the maximum value attainable 
by $EP^{(0)}_{\Gamma_1}$  in the perturbative regime.
In fact it is possible to reproduce, with a good approximation, the results of our simulations by using this probability.
This can be done by calculating the mode entanglement of the following state:
\be
\ket {\phi}=\sqrt{1 - P} \ket {\psi_{in}} + \sqrt P \ket {I}
\ee
where 
\be
P=P(\varepsilon,\tau)=\frac{4|W|^2}{4|W|^2 + (E_1 - E_2)^2}
\ee
is the maximum attainable value of $P_{\ket {\psi_{in}} \rightarrow \ket{I}}(t)$ according to the Rabi's formula.
In the case of the system $\Gamma_0$, the state $\ket {I}$ is given by a superposition of states for which $n_0 = 0,2$. The only contribution to the terms $\rho_0^{(0)}, \rho_2^{(0)}$ of the reduced density matrix of the mode $0$ are given by these states. This means, as described in the previous paragraphs, that the higher are the values of these terms the "flatter" is the probability distribution $(\rho^{(0)}_0,\rho^{(0)}_1,\rho^{(0)}_2,\rho^{(0)}_3,)$ and consequently the higher is the value of $EP^{(0)}_{\Gamma_j}$. For this reason the state $\ket {\phi}$ can be considered as an example of state in which the  entanglement is maximum, and we can compute the entangling power for any rooted graph, and for all the values of 
$\varepsilon$ in the perturbative regime as $ EP^{(0)}_{\Gamma_j}= -Tr(\rho^{(0)} \log_2\rho^{(0)})$
where $\rho^{(0)}  =  \mbox{Tr}_{V_{\backslash 0}}(\ket{\phi}\bra{\phi})$.

By means of the two levels picture we now have the opportunity to see from another  perspective the difference between bipartite and non-bipartite graphs.
This result is essential to explain the difference in the behaviour of 
$EP^{(0)}_{\Gamma_1}$ described by equations (6) and (7).
Since both $\ket{\psi_{in}}$ and $\ket{I}$ are eigenvectors of $H_\varepsilon$, 
if $\sandwich{I}{H_\tau}{I} = 0$, the difference $(E_1 - E_2)$ does not depend 
on $\tau$ and is linear in $\varepsilon$;  thus the sign of the latter does not 
have any effect on $P$. This is always true for all bipartite graphs.To demonstrate that 
for all bipartite graphs $\sandwich{I}{H_\tau}{I} = 0$, we first observe that, 
since  $\ket{I} \propto H_\tau \ket{\psi_{in}}$, we can always write
\be
\sandwich{I}{H_\tau}{I} \propto \sandwich{\psi_{in}}{H_\tau^3}{\psi_{in}} = 
\sandwich{111}{H_\tau^3}{111}.
\ee
We can then make use of the transformation $P=\exp(-\pi \sum_{j \in A} n_j)$ 
previously described to demonstrate that for any bipartite graph 
$\sandwich{111}{H_\tau^3}{111} = 0$. In order to do this we calculate the 
expectation value $\sandwich{111}{P^\dagger H_\tau^3 P}{111}$ in two equivalent 
ways; in the first we let act $P (P^\dagger)$ on $\ket{111} (\bra{111})$ to 
obtain:
\[
 (\bra{111} P^\dagger) H_\tau^3 (P \ket{111}) = \sandwich{111}{ H_\tau^3}{111}.  
\]
In the second one we use the fact that for any bipartite graph $P^\dagger H_\tau 
P = - H_\tau$; this allows to write 
\[
\sandwich{111}{P^\dagger H_\tau^3 P}{111} = -\sandwich{111}{ H_\tau^3}{111}.  
\]
This implies that for any bipartite graph
\be
\sandwich{I}{H_\tau}{I} = \sandwich{111}{H_\tau^3}{111} = 0.
\ee

The last result is, in general, no longer true when the graph is non-bipartite. 
In fact, for $\Gamma_1$, $\sandwich{I}{H_\tau}{I}$ is constant and proportional 
to $\tau$. This is essential to explain the difference in the behaviour of 
$EP^{(0)}_{\Gamma_1}$ in correspondance of an attractive or a repulsive 
self-interaction. In fact, the presence  in $E_2$ of the term proportional to 
$\tau$ renders the difference $(E_1 -E_2)$ sentive to the sign of $\varepsilon$.
\[
(E_1 -E_2)=-(2\varepsilon +3\tau).
\]
This implies that
\[
P(-|\varepsilon|) > P(+|\varepsilon|),
\]
and consequently, as one can see by calculating  $EP^{(0)}_{\Gamma_1}$ from the state $\ket {\phi}$,
\[
EP^{(0)}_{\Gamma_1}(\ket{111},-
|\varepsilon|)>EP^{(0)}_{\Gamma_1}(\ket{111},+|\varepsilon|).
\]

Since the resuts $(12)$ and its consequence $(13)$ hold also for 
$\ket{\psi_{in}}=\ket{300}$ we can use the same arguments to give a 
justification of $(1)$. Here simple calculations give:
\[
E_1 -E_2= 4\varepsilon -\tau.
\]
The role of $\varepsilon$ is then opposite with respect to the previous case: 
\[
P(+|\varepsilon|) > P(-|\varepsilon|),
\]
and consequently, in the perturbative limit 
\[
EP^{(0)}_{\Gamma_1}(\ket{300},+|\varepsilon|) \geq 
EP^{(0)}_{\Gamma_1}(\ket{300},-|\varepsilon|).
\]

\begin{figure}
\putfig{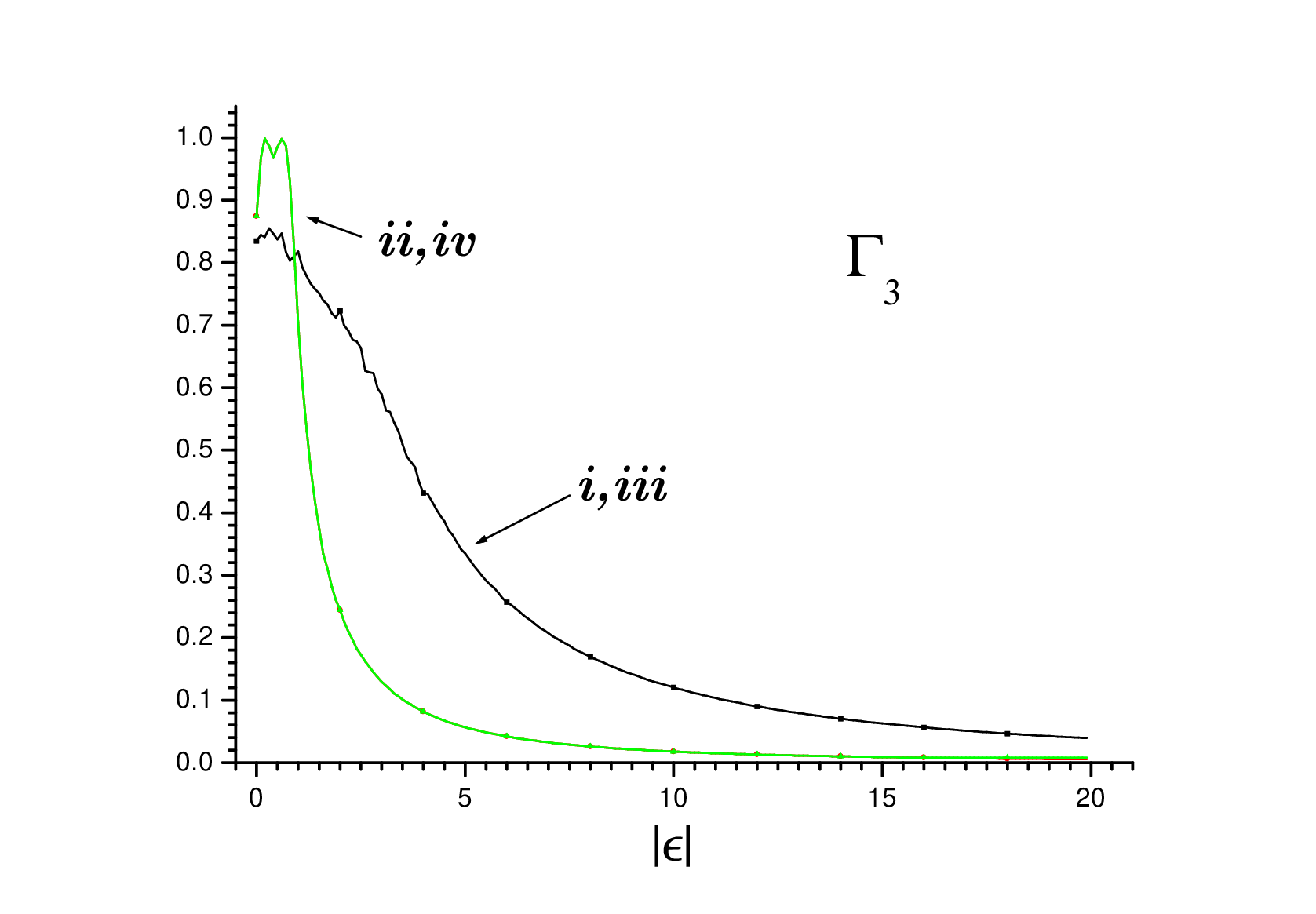}{10}
\caption{ Plot of $EP_{\Gamma_{1}}^0 (|\varepsilon|)$ for 
$\ket{\psi_{in}}=\ket{1111} \ (i \mbox{ and } iii)$ and for 
$\ket{\psi_{in}}=\ket{4000}\ (ii \mbox{ and } iv) $; the curves $i \mbox{ and } 
ii$ refer to $|\varepsilon|=\varepsilon$, while the curves $iii \mbox{ and } vi$ 
refer to $|\varepsilon|=-\varepsilon$.}
\label{array}
\end{figure}

\begin{figure}
\putfig{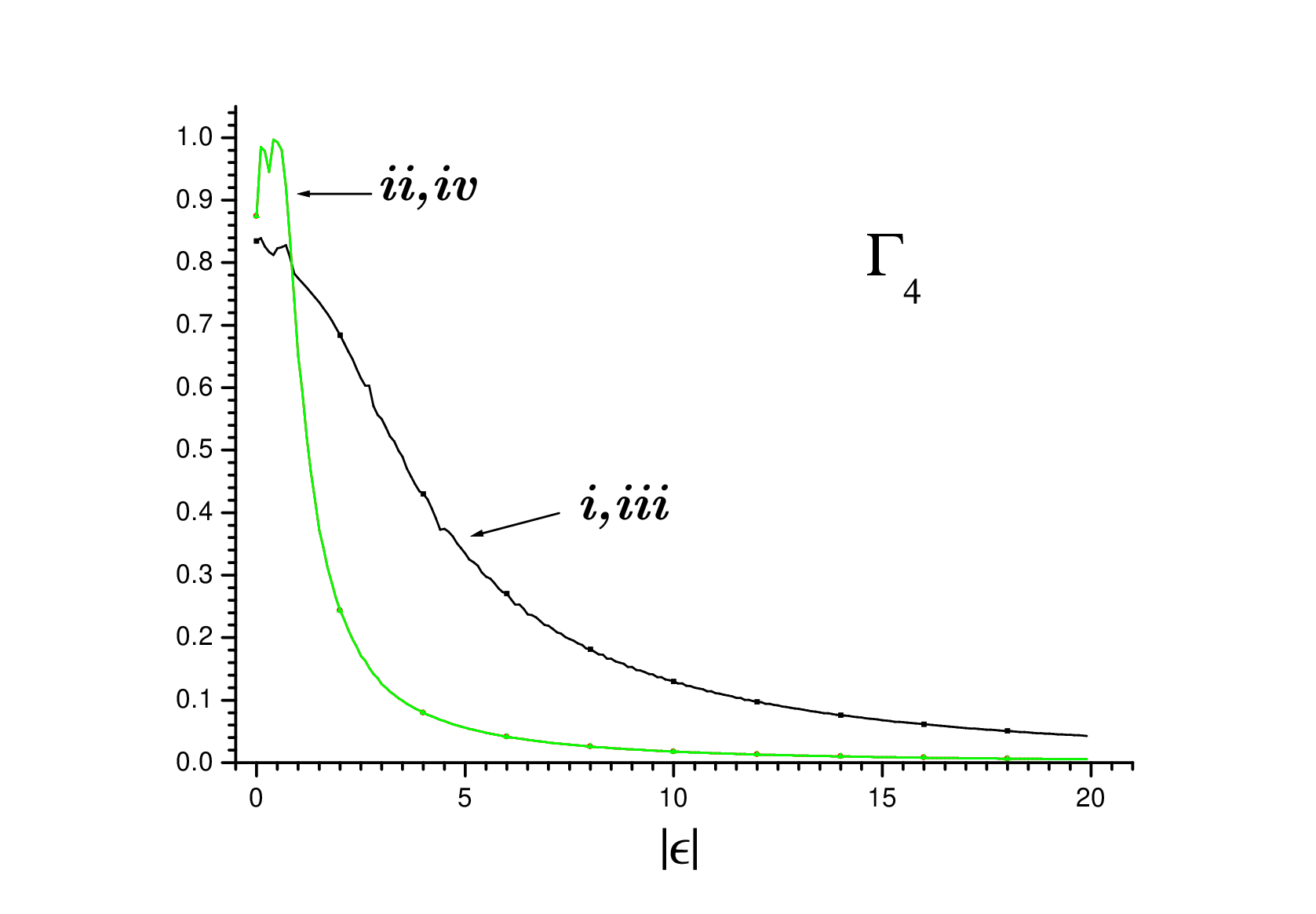}{10}
\caption{ Plot of $EP_{\Gamma_{1}}^0 (|\varepsilon|)$ for 
$\ket{\psi_{in}}=\ket{1111} \ (i \mbox{ and } iii)$ and for 
$\ket{\psi_{in}}=\ket{4000}\ (ii \mbox{ and } iv) $; the curves $i \mbox{ and } 
ii$ refer to $|\varepsilon|=\varepsilon$, while the curves $iii \mbox{ and } vi$ 
refer to $|\varepsilon|=-\varepsilon$.}
\label{array}
\end{figure}

\begin{figure}
\putfig{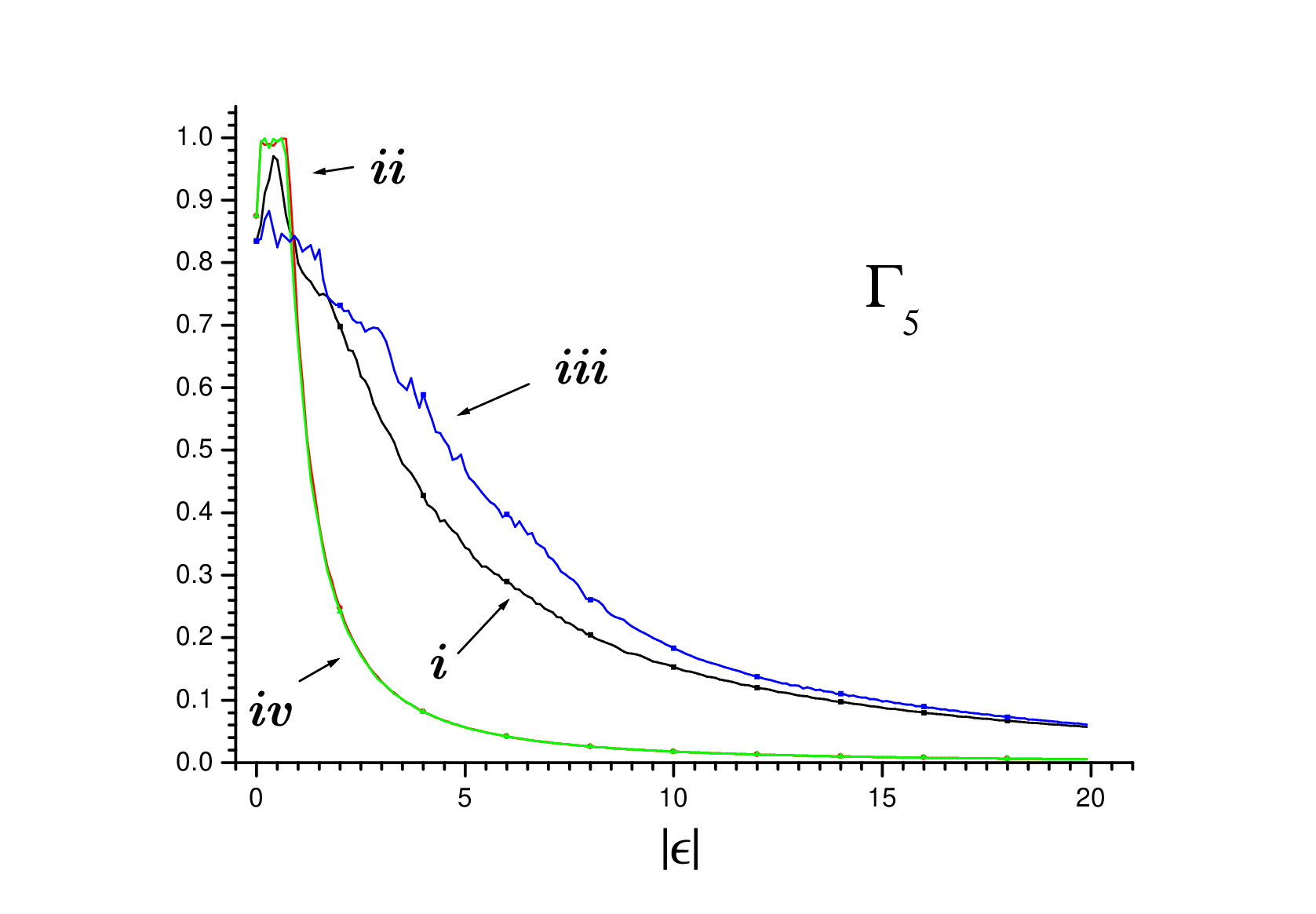}{10}
\caption{ Plot of $EP_{\Gamma_{1}}^0 (|\varepsilon|)$ for 
$\ket{\psi_{in}}=\ket{1111} \ (i \mbox{ and } iii)$ and for 
$\ket{\psi_{in}}=\ket{4000}\ (ii \mbox{ and } iv) $; the curves $i \mbox{ and } 
ii$ refer to $|\varepsilon|=\varepsilon$, while the curves $iii \mbox{ and } vi$ 
refer to $|\varepsilon|=-\varepsilon$.}
\label{array}
\end{figure}

\begin{figure}
\putfig{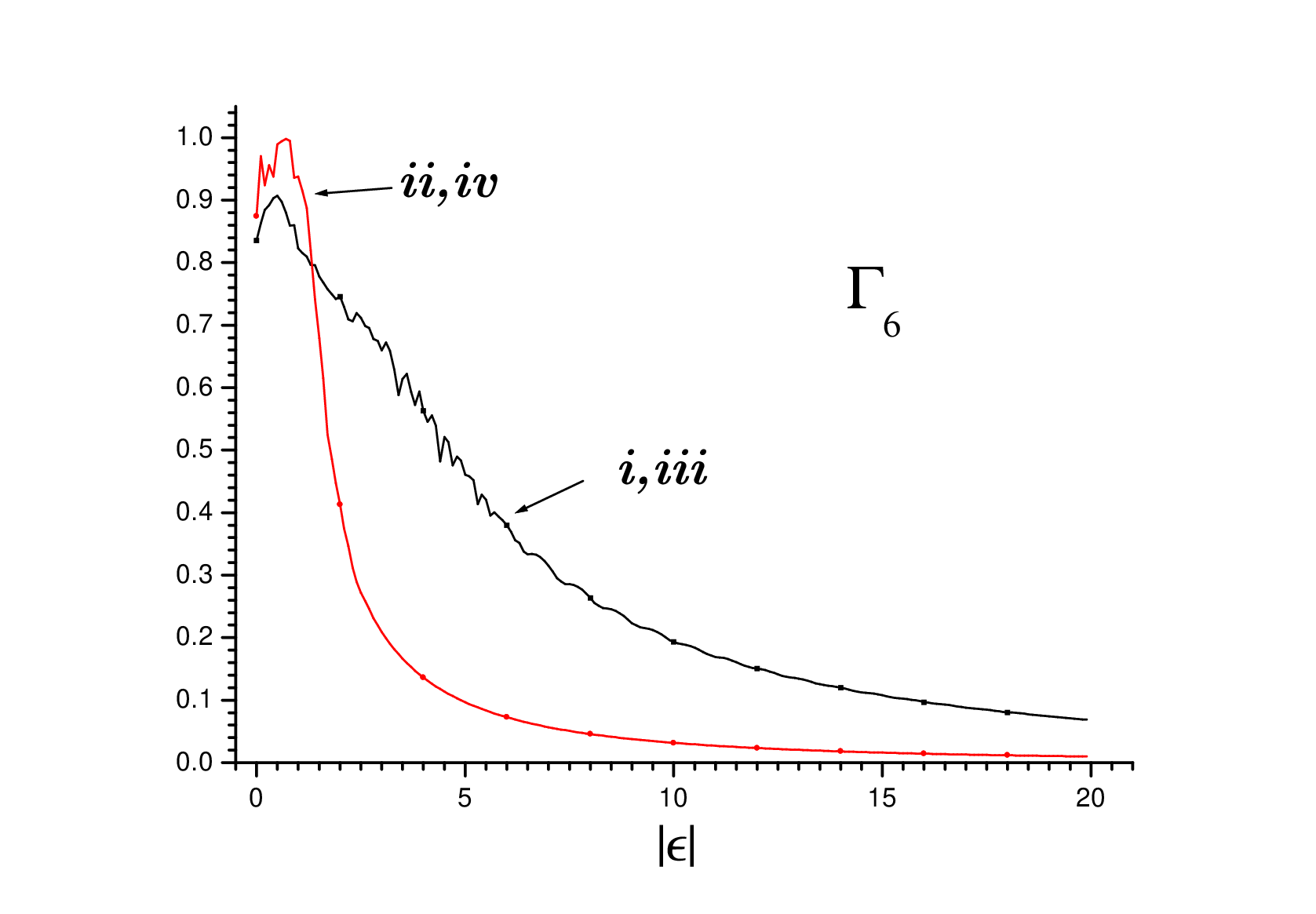}{10}
\caption{ Plot of $EP_{\Gamma_{1}}^0 (|\varepsilon|)$ for 
$\ket{\psi_{in}}=\ket{1111} \ (i \mbox{ and } iii)$ and for 
$\ket{\psi_{in}}=\ket{4000}\ (ii \mbox{ and } iv) $; the curves $i \mbox{ and } 
ii$ refer to $|\varepsilon|=\varepsilon$, while the curves $iii \mbox{ and } vi$ 
refer to $|\varepsilon|=-\varepsilon$.}
\label{array}
\end{figure}

\begin{figure}
\putfig{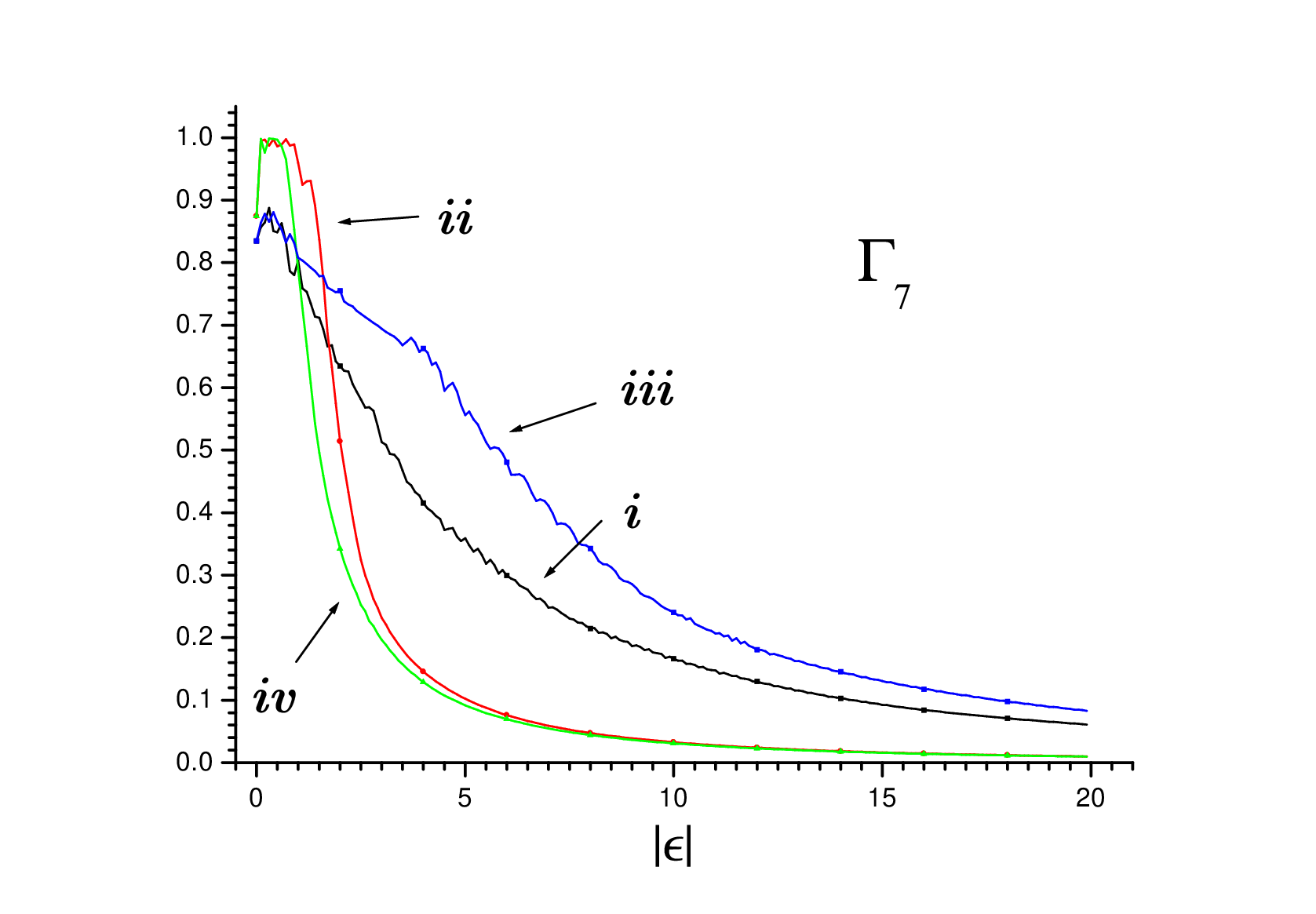}{10}
\caption{ Plot of $EP_{\Gamma_{1}}^0 (|\varepsilon|)$ for 
$\ket{\psi_{in}}=\ket{1111} \ (i \mbox{ and } iii)$ and for 
$\ket{\psi_{in}}=\ket{4000}\ (ii \mbox{ and } iv) $; the curves $i \mbox{ and } 
ii$ refer to $|\varepsilon|=\varepsilon$, while the curves $iii \mbox{ and } vi$ 
refer to $|\varepsilon|=-\varepsilon$.}
\label{array}
\end{figure}

\begin{figure}
\putfig{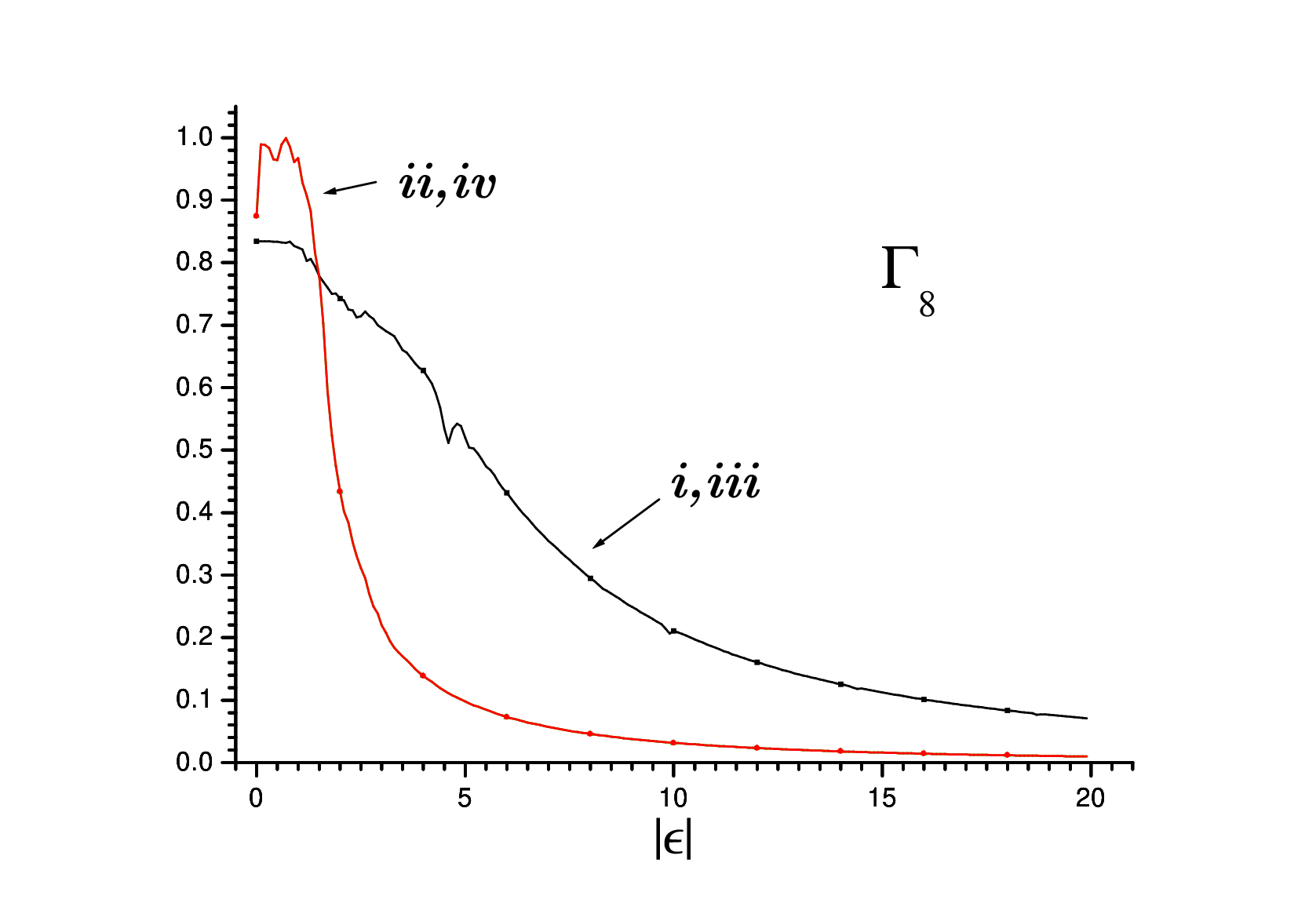}{10}
\caption{ Plot of $EP_{\Gamma_{1}}^0 (|\varepsilon|)$ for 
$\ket{\psi_{in}}=\ket{1111} \ (i \mbox{ and } iii)$ and for 
$\ket{\psi_{in}}=\ket{4000}\ (ii \mbox{ and } iv) $; the curves $i \mbox{ and } 
ii$ refer to $|\varepsilon|=\varepsilon$, while the curves $iii \mbox{ and } vi$ 
refer to $|\varepsilon|=-\varepsilon$.}
\label{array}
\end{figure}

\begin{figure}
\putfig{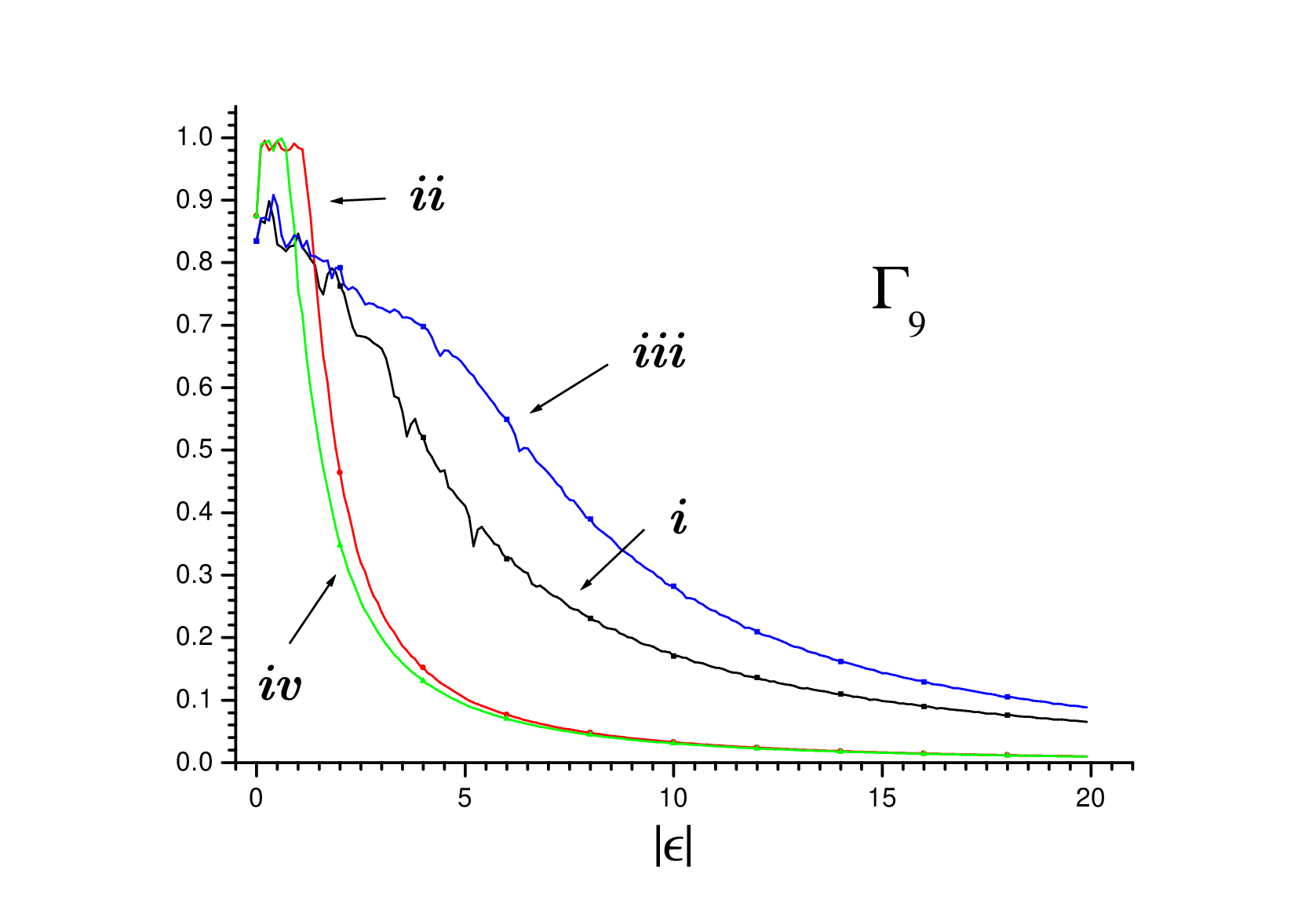}{10}
\caption{ Plot of $EP_{\Gamma_{1}}^0 (|\varepsilon|)$ for 
$\ket{\psi_{in}}=\ket{1111} \ (i \mbox{ and } iii)$ and for 
$\ket{\psi_{in}}=\ket{4000}\ (ii \mbox{ and } iv) $; the curves $i \mbox{ and } 
ii$ refer to $|\varepsilon|=\varepsilon$, while the curves $iii \mbox{ and } vi$ 
refer to $|\varepsilon|=-\varepsilon$.}
\label{array}
\end{figure}

\begin{figure}
\putfig{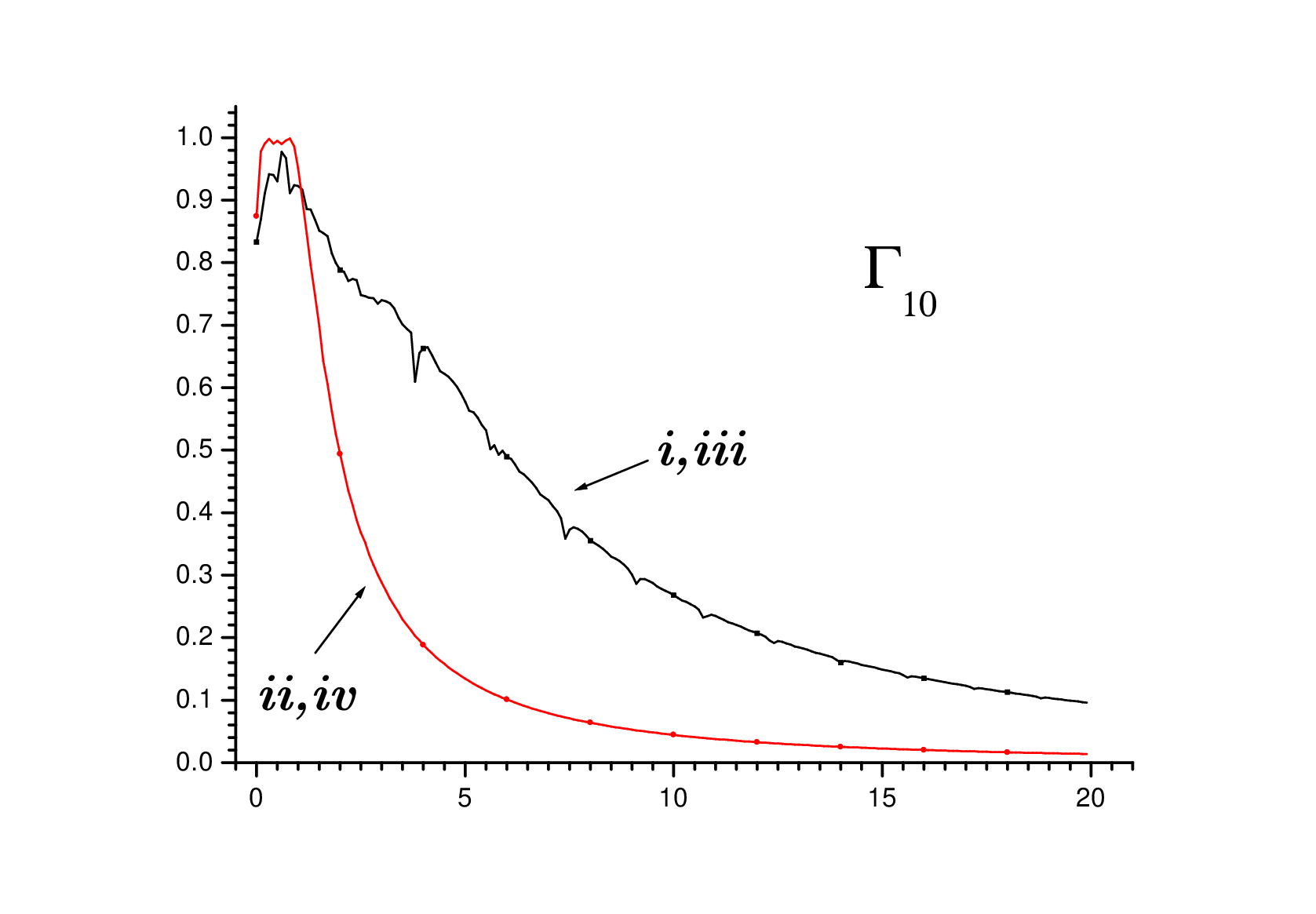}{10}
\caption{ Plot of $EP_{\Gamma_{1}}^0 (|\varepsilon|)$ for 
$\ket{\psi_{in}}=\ket{1111} \ (i \mbox{ and } iii)$ and for 
$\ket{\psi_{in}}=\ket{4000}\ (ii \mbox{ and } iv) $; the curves $i \mbox{ and } 
ii$ refer to $|\varepsilon|=\varepsilon$, while the curves $iii \mbox{ and } vi$ 
refer to $|\varepsilon|=-\varepsilon$.}
\label{array}
\end{figure}

\begin{figure}
\putfig{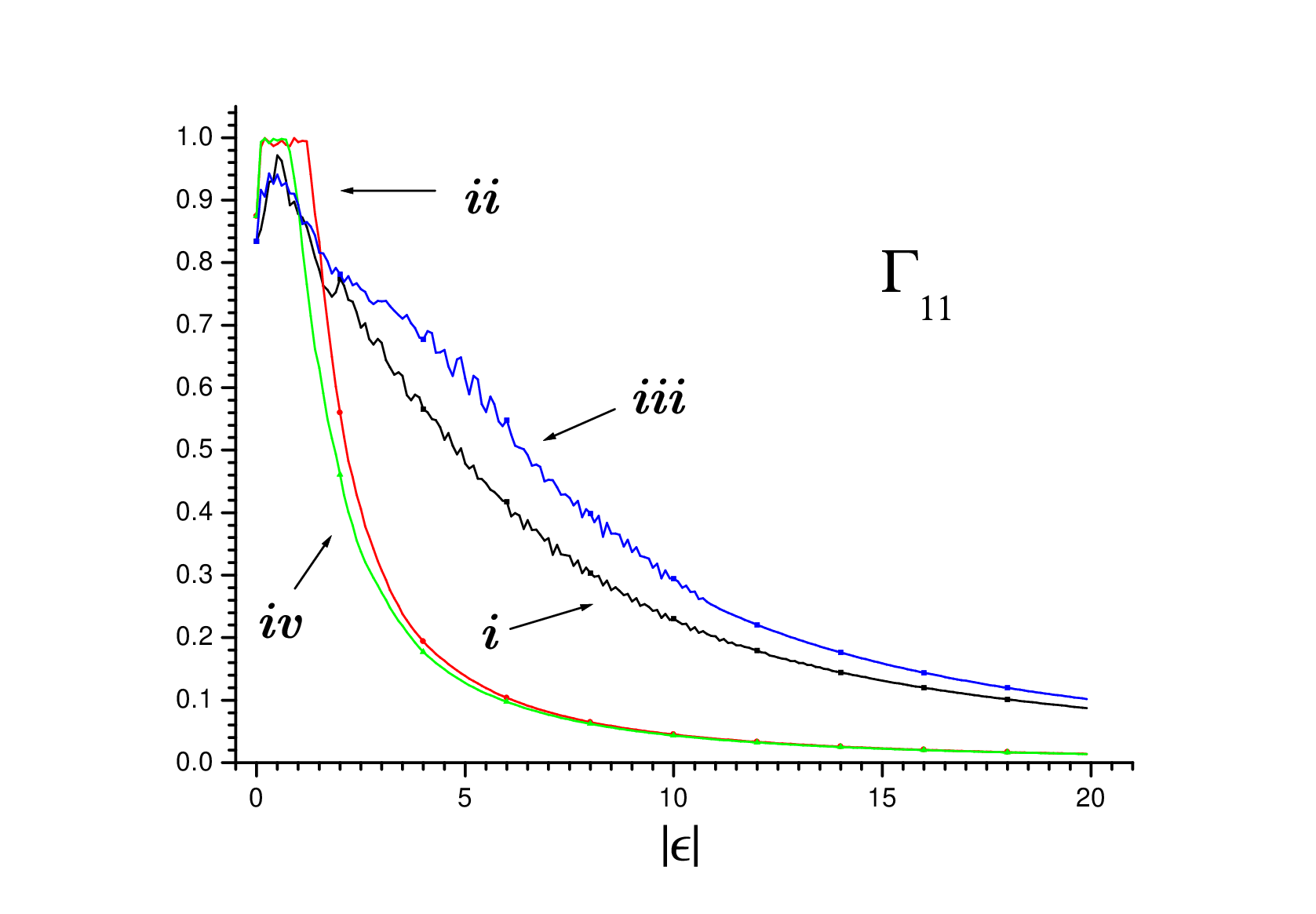}{10}
\caption{ Plot of $EP_{\Gamma_{1}}^0 (|\varepsilon|)$ for 
$\ket{\psi_{in}}=\ket{1111} \ (i \mbox{ and } iii)$ and for 
$\ket{\psi_{in}}=\ket{4000}\ (ii \mbox{ and } iv) $; the curves $i \mbox{ and } 
ii$ refer to $|\varepsilon|=\varepsilon$, while the curves $iii \mbox{ and } vi$ 
refer to $|\varepsilon|=-\varepsilon$.}
\label{array}
\end{figure}

\begin{figure}
\putfig{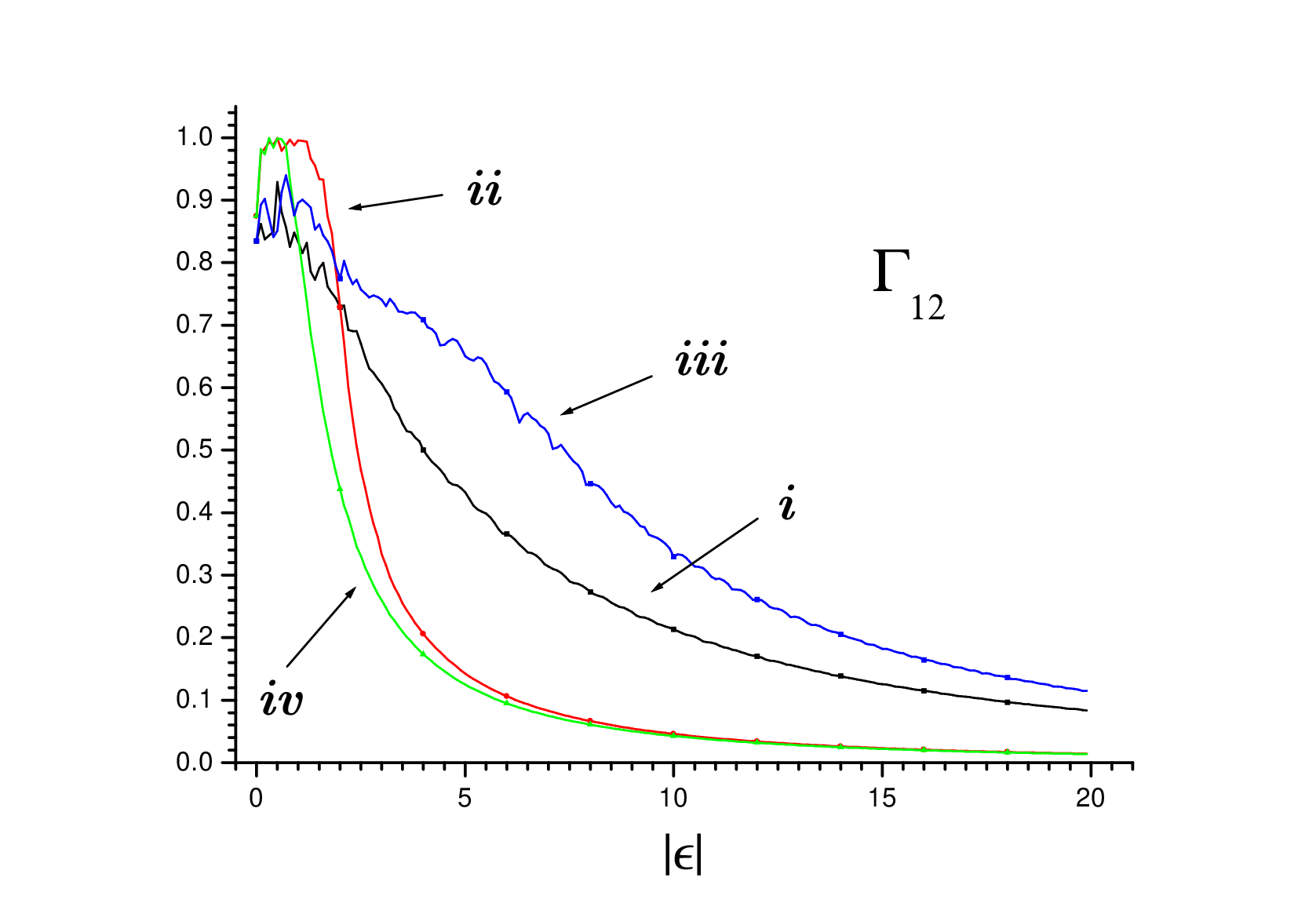}{10}
\caption{ Plot of $EP_{\Gamma_{1}}^0 (|\varepsilon|)$ for 
$\ket{\psi_{in}}=\ket{1111} \ (i \mbox{ and } iii)$ and for 
$\ket{\psi_{in}}=\ket{4000}\ (ii \mbox{ and } iv) $; the curves $i \mbox{ and } 
ii$ refer to $|\varepsilon|=\varepsilon$, while the curves $iii \mbox{ and } vi$ 
refer to $|\varepsilon|=-\varepsilon$.}
\label{array}
\end{figure}

\begin{figure}
\putfig{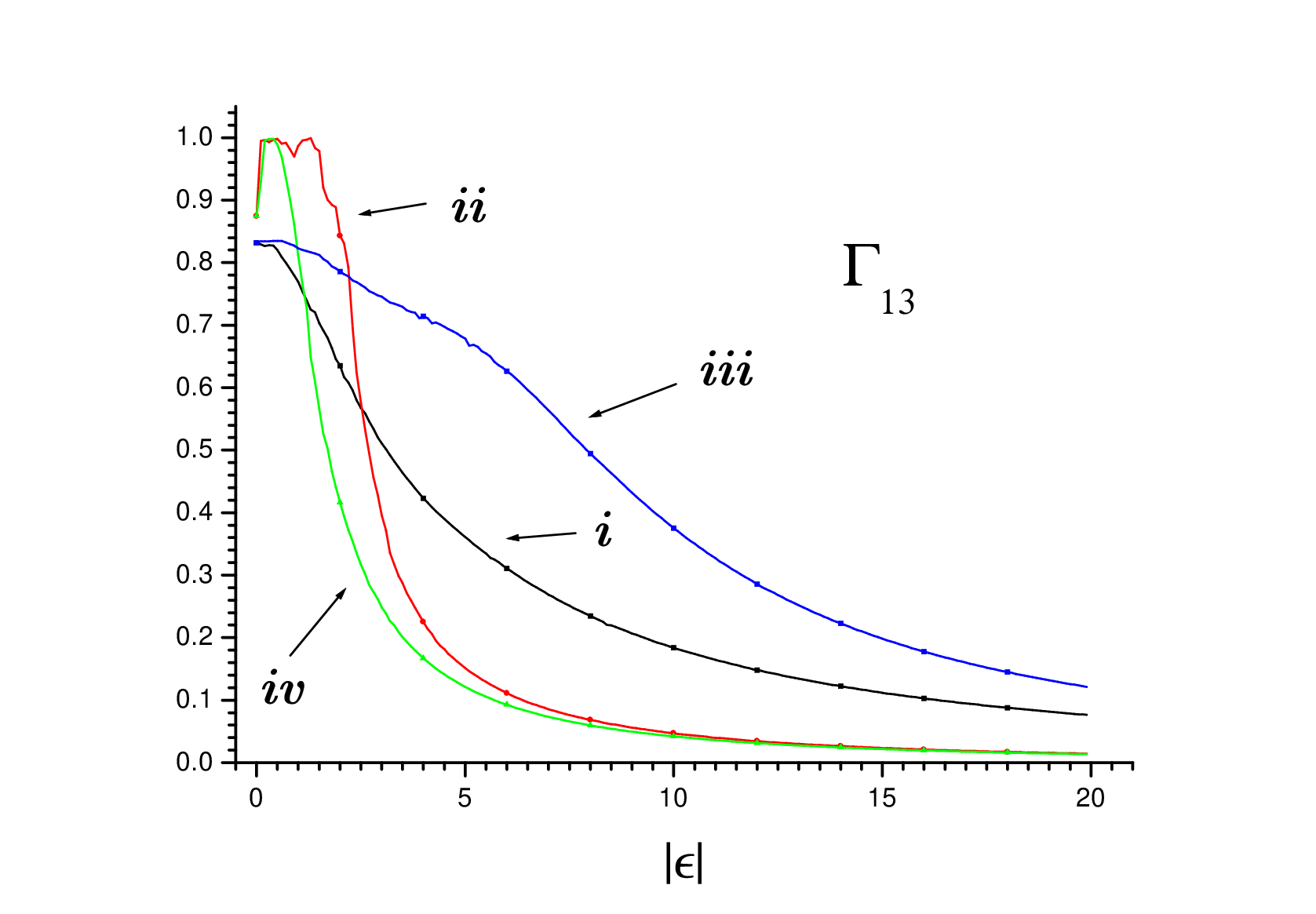}{10}
\caption{ Plot of $EP_{\Gamma_{1}}^0 (|\varepsilon|)$ for 
$\ket{\psi_{in}}=\ket{1111} \ (i \mbox{ and } iii)$ and for 
$\ket{\psi_{in}}=\ket{4000}\ (ii \mbox{ and } iv) $; the curves $i \mbox{ and } 
ii$ refer to $|\varepsilon|=\varepsilon$, while the curves $iii \mbox{ and } vi$ 
refer to $|\varepsilon|=-\varepsilon$.}
\label{array}
\end{figure}

\subsection{N=L=4}
In this subsection we illustrate the results of our simulations for the case 
$N=L=4$. In the following we refer to the figures 6-16. For each graph 
$\Gamma_j, j\in\{3,4,..,13\}$ we have computed $EP^{(0)}_{\Gamma_j}$ as a function 
of $\varepsilon $, using  $\triangle \varepsilon = 0.1$. For each value of 
$\varepsilon$ we evolved the system from $t=0$ to $T=15$ (arbitrary 
units).
The curves $(i)$ and $(iii)$ in each graph are the results of the 
simulations for $\ket{\psi_{in}}=\ket{1111}$. The curves
$(ii)$ and $(iv)$ refer to the simulations in which the initial state is 
$\ket{\psi_{in}}=\ket{4000}$. In the $(i)$ and $(ii)$ curves $\varepsilon$ varies in the interval $[0,\varepsilon_{max}]$, while in the 
$(iii)$ and  $(iv)$ curves $\varepsilon\in [-\varepsilon_{max},0]$; for all the 
different cases $\varepsilon_{max}=+20$ and the value of $EP^{(0)}_{\Gamma_j}$ has been plotted in function 
of $|\varepsilon|$.

\vspace{0.2cm}

As we did in the previous section we start by observing that when $\varepsilon = 
0$ the entagling power of the evolution operator $U_{\Gamma_j}$ does not depend 
on the graph $\Gamma_j$ but only on the chosen initial state. In fact 
$EP^{(0)}_{\Gamma_j} \cong 0.83\  \mbox{, for any } j $, in the case 
$\ket{\psi_{in}}=\ket{1111}$, while $EP^{(0)}_{\Gamma_j} \cong 0.87 \ \mbox{, for any } j$, when $\ket{\psi_{in}}=\ket{4000}$. We see here a first difference with case 
$N=L=3$: the entagling capability of the $U_{\Gamma_j}$'s , when $\varepsilon = 
0$, is enhanced if the system starts with all particles localized in the mode $0$.

\vspace{0.2cm}

We now turn to analize the maximum values reached by $EP^{(0)}_{\Gamma_j}$ for 
the different sistems. The maximum theoretical value of the mode entanglement 
$(\log_2 5)$ is always  reached $\forall \ \Gamma_j$ when the initial state is 
$\ket{4000}$. In fact, for an appropriate choice of the ratio $|\varepsilon| / 
\tau$, regardless of the sign of the self-interaction, 
$ EP^{(0)}_{\Gamma_j}(\ket{4000}) > 0.99$ . In contrast with the 
case $N=L=3$,  none of the systems reaches this value when the initial state is 
the uniformly occupied one; the highest values of 
$EP^{(0)}_{\Gamma_j}(\ket{1111},)$ are $\gtrsim 0.97$ and they are 
reached by the systems $\Gamma_5, \Gamma_{11}$ ( only for $\varepsilon > 0$), and $\Gamma_{10}$.

When $\ket{\psi_{in}}=\ket{4000}$, $EP^{(0)}_{\Gamma_j}$ increases when 
$|\varepsilon|$ starts to differ from $0$ and passes from $0.87$ to values $> 
0.92$ in correspondance of $|\varepsilon| = 0.1$; then the value of 
$EP^{(0)}_{\Gamma_j}$ remains very high for $|\varepsilon| < \varepsilon'$, 
where $\varepsilon'>0$ differs for the various cases. We can then say that, in 
this interval of values the self-interaction  between the particles, either 
repulsive or attractive, enhances the capability of the $U_{\Gamma_j}$'s to 
entangle the mode $0$ with the rest of the system. 

In the case $\ket{\psi_{in}}=\ket{1111}$, as mentioned above, the maximum 
theoretical value of the mode entanglement is never reached and the effect of 
turning on the self-interaction depends on the system and on the sign of the 
self-interaction. However, the enhancement of the entangling capability of the 
$U_{\Gamma_j}$'s, when present, is found in correspondance of values of 
$|\varepsilon|$ which are smaller or of the same order of $\tau (=1)$.

\vspace{0.2cm}

By observing the figures 6-16 we can see that also for the case $N=L=4$ we can 
distinguish between two different behaviours of $EP^{(0)}_{\Gamma_j}$ with 
respect to the sign change of $\varepsilon$. For $j \in \{3,4,6,8,10 \}$ we 
have that for both the intial states
\be
EP^{(0)}_{\Gamma_j}(\varepsilon)=EP^{(0)}_{\Gamma_j}(-\varepsilon)\,\   \forall 
\varepsilon \in [0,20], 
\ee
that is the curves $i$ and $ii$ perfectly overlap with the curves $iii$ and $iv$ 
respectively. On the other hand for 
$j \in \{5,7,9,11,12,13 \}$ this is no longer true. 

This fact can be explained using the same arguments we used in the 
previous section. In fact the $\Gamma_j, j=3,..13$, can be distinguished in {\it bi-partite} and {\it non-bipartite} graphs. For the bipartite ones we can apply to the 
related Hamiltonians the unitary transformation $(4)$ in order to get equation 
$(3)$. The latter and the subsequent reasonings hold for any $N=L$ and thus 
equation $(5)$ implies that  $\forall j \in \{3,4,6,8,10 \}$, i.e. for all 
bipartite graphs equation $(14)$ must hold.

\vspace{0.2cm}

We now turn to examine the behaviour of $EP^{(0)}_{\Gamma_j}$ in the pertubative 
regime. We first observe that for both the initial states  $EP^{(0)}_{\Gamma_j}$ 
rapidly decreses when $|\varepsilon|$ is significantly greater than $1 (=\tau)$. 
This happens for the same reasons we described for the case $N=L=3$. The 
Hamiltonian, $\forall \Gamma_j$, is composed of an unperturbed diagonal term 
$H_\varepsilon$, the self-interaction one, and a perturbation term 
given by the hopping matrix $H_\tau$. Both the initial states of our computation 
are eigenvectors of the unperturbed term and correspond to the eigenvalues 
$4\varepsilon$ ($\ket{1111}$) and $16\varepsilon$ ($\ket{4000}$). As 
$|\varepsilon|$ increses, the perturbation becomes small and the only processes 
that become relevant for the evolution are the first order ones.
At first order, the state $\ket{\psi_{in}}$ can be coupled  by the perturbation 
to the states $\ket{\vec n'} \in \chi'_{ \ket{ \psi_{in} }}$. This implies that 
for  $\ket{\psi_{in}}=\ket{1111}$ we have that $\rho^{(0)}_3, \rho^{(0)}_4 \ll 
\rho^{(0)}_0, \rho^{(0)}_2 \ll \rho^{(0)}_1$, while for 
$\ket{\psi_{in}}=\ket{4000}$ we have that $\rho^{(0)}_0, \rho^{(0)}_1, 
\rho^{(0)}_2 \ll \rho^{(0)}_3 \ll \rho^{(0)}_4$. The probability distribution
$(\rho^{(0)}_0, \rho^{(0)}_1, \rho^{(0)}_2 ,\rho^{(0)}_3, \rho^{(0)}_4)$ is 
higly unbalanced in both cases, and this results in the decreasing behaviour of 
$EP^{(0)}_{\Gamma_j}$, but it is more unbalanced in the case 
$\ket{\psi_{in}}=\ket{4000}, \forall j, $ and this is a first explanation of the 
fact that in the perturbative regime
\be
EP^{(0)}_{\Gamma_j}(\ket{1111}) > EP^{(0)}_{\Gamma_j}(\ket{4000}), \ \forall j.
\ee
A further explanation for the last relation is given, just as for the case 
$N=L=3$, by the fact that the strenght of the first order coupling between 
$\ket{\psi_{in}}$ 
and the states belonging to $\chi'_{ \ket{ \psi_{in} }}$ is higher in the case 
$\ket{\psi_{in}}=\ket{1111},(\tau / \triangle E = 1/(2|\varepsilon|),$ then in 
the case $\ket{\psi_{in}}=\ket{4000},(\tau / \triangle E = 1/(6|\varepsilon|)$.
As explained in the previous section this affects the highest reachable value 
for the probability related to the states belonging to in $\chi'_{ \ket{ 
\psi_{in} }}$ and consequently affects the values reached by  
$EP^{(0)}_{\Gamma_j}$ in the way described by $(15)$.

\vspace{0.2cm}

We finally come to the description of the role of the sign of the 
self-interaction when $|\varepsilon|$ is significantly greater than $1 (=\tau)$  
in the case of the non-bipartite graphs $\Gamma_j, \ j\in\{5,7,9,11,12,13 \}$. 
The phenomenology is analogous to the one found for the case $N=L=3$. For 
$\ket{\psi_{in}}=\ket{4000}$ a repulsive (positive) self-interaction enhances the 
entangling capability of $U_{\Gamma_j}(t)$, that is for a large part of the 
interval $[0,|\varepsilon_{max}|]$
\be
EP^{(0)}_{\Gamma_j}(\ket{4000},+|\varepsilon|) \geq 
EP^{(0)}_{\Gamma_j}(\ket{4000},-|\varepsilon|).
\ee
In contrast, for $\ket{\psi_{in}}=\ket{1111}$ it is the attractive (negative) 
self-interaction that for a large part of the interval $[0,|\varepsilon_{max}|]$ 
enhances the entangling power of $U_{\Gamma_j}(t)$:
\be
EP^{(0)}_{\Gamma_j}(\ket{1111},-|\varepsilon|) > 
EP^{(0)}_{\Gamma_j}(\ket{1111},+|\varepsilon|).
\ee
This behaviour can be understood by means of the same arguments exposed for the 
case $N=L=3$ that we briefly recall here. When $|\varepsilon|$ is significantly 
greater than $1 (=\tau)$, we can again use a two-levels picture to describe our 
systems. The first level is $\ket{\psi_{in}}$, while the second $\ket{I}$ is 
given by a uniform superposition of the $q = |\chi'_{ \ket{ \psi_{in} } }|$ 
degenerate states that are coupled to  $\ket{\psi_{in}}$ by first order 
processes, see eq. (8). The two-levels Hamiltonian is again given by (see  
formula) and the probability $P_{\ket {\psi_{in}} \rightarrow \ket{I}}(t)$ is 
expressed by the Rabi's formula. As we have seen for the case $N=L=3$ with the introduction of the state $\ket{\phi}$, the 
maximum attainable value of this probability $P=P(\varepsilon,\tau)$ is directly related to the value 
of $EP^{(0)}_{\Gamma_j}$, that is the higher the former the higher the latter. 
In this context it is again the term $(E_1 - E_2)$ that is crucial to give an 
explanation of the equations (16) and (17). In fact, equation (13) holds also 
for the case $N=L=4$ and for the intial states $\ket{1111} \mbox{ and } 
\ket{4000}$. Thus, for the bipartite graphs, we have that 
$\sandwich{I}{H_\tau}{I} = 0$, the term $(E_1 - E_2)$ is independent of $\tau$ 
and depends linearly on $\varepsilon$. This implies that the probability 
$P_{\ket {\psi_{in}} \rightarrow \ket{I}}(t)$ does not depend on the sign of 
$\varepsilon$  for the bipartite graphs $\Gamma_j, \ j\in\{3,4,6,8,10\}$. 

In the case of non-bipartite graphs this is no longer true and the dependance of 
$(E_1 - E_2)$ on $\tau$ give rise to a sensitivity to the sign of $\varepsilon$ 
of $P=P(\varepsilon,\tau)$. Hence the different behaviour of $EP^{(0)}_{\Gamma_j}$ 
described by equations (16) and (17).

\begin{figure}
\putfig{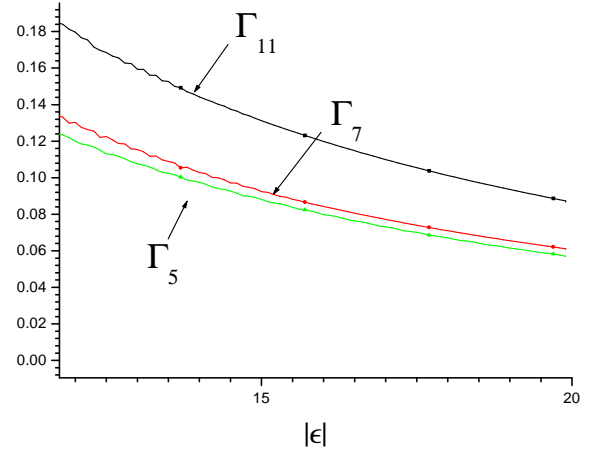}{10}
\caption{ Plot of $EP_{\Gamma_{j}}^0 (|\varepsilon|)$ for $j=5,7,11$;
$\ket{\psi_{in}}=\ket{1111}$ and  $|\varepsilon|=\varepsilon$.}
\label{array}
\end{figure}

\begin{figure}
\putfig{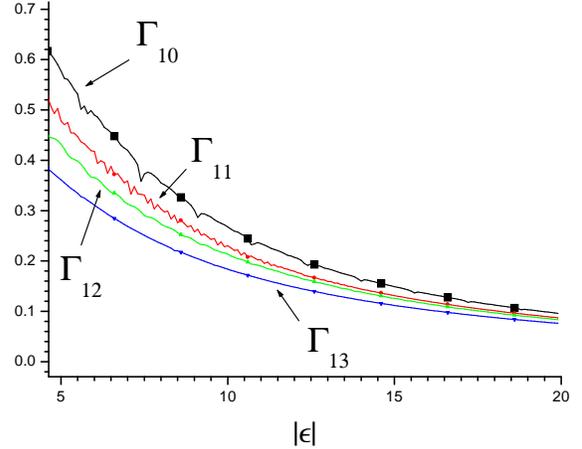}{10}
\caption{ Plot of $EP_{\Gamma_{j}}^0 (|\varepsilon|)$ for $j=10,11,12,13$;
$\ket{\psi_{in}}=\ket{1111}$ and  $|\varepsilon|=\varepsilon$.}
\label{array}
\end{figure}

\begin{figure}
\putfig{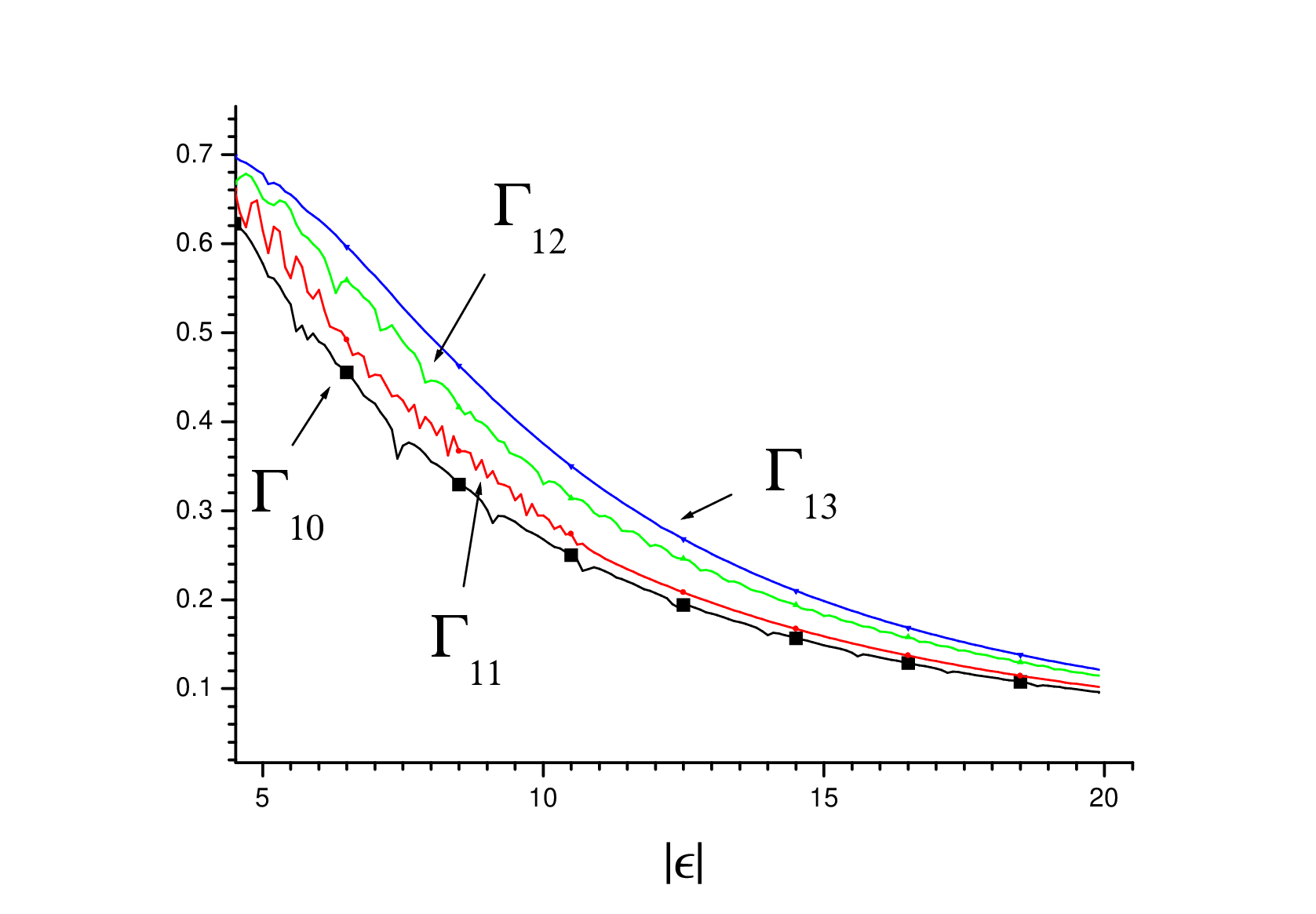}{10}
\caption{ Plot of $EP_{\Gamma_{j}}^0 (|\varepsilon|)$ for $j=10,11,12,13$;
$\ket{\psi_{in}}=\ket{1111}$ and  $|\varepsilon|=-\varepsilon$.}
\label{array}
\end{figure}

\subsection{Graph Properties and Entangling Power in the Pertubative Regime}

In the last part of the paper we want to deepen our analysis in the perturbative regime in order to describe the existing relations among the behaviour of $EP^{(0)}_{\Gamma_j}$ for the different $\Gamma_j$. 
As we have seen in the previous paragraphs, in the perturbative regime, this behaviour can be described by using for our systems a two levels approach. In this picture, the maximum value attainable for $EP^{(0)}_{\Gamma_j}$ is directly linked with $P$, i.e. the maximum of the probability $P_{\ket {\psi_{in}} \rightarrow \ket{I}}(t)$, see equations (8)-(11).
It is possible to reproduce, with a good approximation, the results of our simulations by using this probability and by calculating the mode entanglement of the state (10), that is:
\be
 EP^{(0)}_{\Gamma_j}= -Tr(\rho^{(0)} \log_2\rho^{(0)})
\ee
where $\rho^{(0)}  =  \mbox{Tr}_{V_{\backslash 0}}(\ket{\phi}\bra{\phi})$.
In order to evaluate (18) we can use analytical expressions for $P$ and for $\rho^{(0)}$ that allow us to link $ EP^{(0)}_{\Gamma_j}$ to some of the topological properties of the graphs. These expressions can be derived for the general case $N=L$.

We first start our analysis by considering $\ket{\psi_{in}}=\otimes_{j=0}^{L-1} \ket{1}_j$;
in this case the expression for P is:
\be
P= \frac{16 \tau^2 k }{16 \tau^2 k+(2\varepsilon  + \tau 18l_3/2k)^2},
\ee
where $16 \tau^2 k = 4|\tau \sqrt{2q}|= 4|W|^2$ is given by the coupling induced by the hopping term $H_{\tau}$ and is proportional to $k=|E|=q/2=|\chi'_{ \ket{ \psi_{in} }}|/2 $, i.e. the number of links of the graph; $2\varepsilon = \sandwich{I}{H_\varepsilon}{I}-\sandwich{\psi_{in}}{H_\varepsilon}{\psi_{in}}$. The term $\tau^2 18 l_3/q = \tau^2 18 l_3/2k = \sandwich{I}{H_\tau}{I}$ depends on $k$ and also on another property of the graph which is the number $l_3$ of closed loops of length $3$ present in the graph (one of such loops can be seen for example in $\Gamma_{11}$ where the loop is given by the links between the sites $0,1$ and $2$). 

We now give a brief explanation of the appearance of the number $l_3$ in the expression in the expression of $P$. As we have already pointed out in section V.A, eq. (12), $\sandwich{I}{H_\tau}{I}$ is proportional to $\sandwich{\psi_{in}}{H_\tau^3}{\psi_{in}}$. The fact that this term differs from zero only if the graph is non bi-partite can be now explained in terms of the presence of loops of length $3$ in the graph. In fact a graph is non bi-partite if and only if it contains at least one of these loops. The presence in the graph of one of such loops results in the presence in $H_\tau^3$ of terms like $T_{i,j,h,k,m,n}= c_i^\dagger c_j c_h^\dagger c_k c_m^\dagger c_n $ such that  
\be
\sandwich{\psi_{in}}{ T_{i,j,h,k,m,n}}{\psi_{in}} \neq 0.
\ee
For example, in the case $\Gamma_{11}$, $ c_0^\dagger c_2 c_2^\dagger c_1 c_1^\dagger c_0 $ is one of such terms.
In order to calculate $\sandwich{I}{H_\tau}{I}$ we first have to compute the number of terms for which equation (20) holds. This can be done, for any loop of length three, as follows. We first observe that the number of possible choices for the index $n$ is $3$, that is the number of vertices involved in the loop. Since each of these is linked to the other $2$ vertices in the loop we have $2$ possible choices for $m$. Once $n$ and $m$ are fixed there are only two kind of terms that satisfy equation (20), that is $a)\ T_{n,h,h,m,m,n}$ and $b)\  T_{h,m,n,h,m,n}$, where $h$ is the third remaining vertex. Consequently, for any  loop  of length three we have $12$ terms that satisfy equation (20). By inspection we see that for the six terms of type $a)\ \sandwich{\psi_{in}}{ T_{n,h,h,m,m,n}}{\psi_{in}}=4$, while for the other six $ \sandwich{\psi_{in}}{ T_{h,m,n,h,m,n}}{\psi_{in}}=2$. This implies that, taking in account of the fixed contribute given by each of the loops of length three present in the graph,  $\sandwich{I}{H_\tau}{I}=18l_3/q$.

\noindent We have then succeded in linking $P$ to the graph properties $k$ and $l_3$.The next step is to explicitly write $EP^{(0)}_{\Gamma_j}$ as the entanglement of the state $\ket {\phi}$; to do this it is necessary to express  $\rho^{(0)}_0,\rho^{(0)}_1$ and $\rho^{(0)}_2$ in terms of the probability $P$.
This can be done in a way that takes in account of the properties of the graph $\Gamma_j$ and of the state $\ket{I}$:
\be
\rho^{(0)}_1 = (1-P)+ \frac{P}{q}2(k-k_0)= 1 - P \frac{k_0}{k}
\ee
\be
\rho^{(0)}_0, \rho^{(0)}_2 = \frac{P}{q}k_0= \frac{Pk_0}{2k},
\ee
where $q = |\chi'_{ \ket{\psi_{in}} }|=2k$; $\ k$ is the number of links in the graph and $k_0$ is the number of links starting from the root vertex. The term $\frac{P}{q}2(k-k_0)$ in (21) takes in account of the $2(k-k_0)$ states in $\ket{I}$ that are coupled by first order processes to $\ket{\psi_{in}}$ and for which $n_0=1$. We can now use the formulas just derived to describe some of the relations existing among $EP^{(0)}_{\Gamma_j}$ for the different rooted graphs in the case $\ket{\psi_{in}}=\otimes_{j=0}^{L-1} \ket{1}_j$.

We first focus on the relation among the values of $EP^{(0)}_{\Gamma_j}$, in the perturbative regime, for rooted graphs that differ only for the choice of the root vertex. As an example we take the rooted graphs $\Gamma_5, \ \Gamma_7$ and $\Gamma_{11}$; in figure (17) we have plotted the results of our simulations for the three graphs in the case $\ket{\psi_{in}}=\ket{1111}$ and $\varepsilon >0$.
From this figure we can see that 
\[
EP^{(0)}_{\Gamma_{11}}>EP^{(0)}_{\Gamma_7} \geq EP^{(0)}_{\Gamma_5}.
\]
That is, the higher the value of $k_0^j$, i.e. of the coordination number of the root vertex of $\Gamma_j$, the higher the value of $EP^{(0)}_{\Gamma_j}$.
This is an example of a feature that holds in general for any value of $N=L$. In fact, suppose to have
\begin{itemize}
\item{}
a set of rooted graph $A=\{ \Gamma_j \}$ that differ one from the other only for the choice of the root vertex; 
\item{}
$\ket{\psi_{in}}=\otimes_{j=0}^{L-1} \ket{1}_j$;
\item{}
a fixed value of $\varepsilon$ ($>0 \mbox{ or } < 0$) in the perturbative regime,
\end{itemize}
then the $\Gamma_j$'s for which $EP^{(0)}_{\Gamma_j}$ is higher are those for which $k_0^j$ is maximum. Furthermore it is possible to say that 
\be
\mbox{\ \ if \ \ } k_0^m > k_0^n \mbox{\ \ then \ \ } EP^{(0)}_{\Gamma_m} > EP^{(0)}_{\Gamma_n}.
\ee
This can be seen from equations (19),(21),(22). In fact, for a given value of $\varepsilon$ in the perturbative regime, $P$ is the same for all the rooted graphs in $A$, since for all them $k$ and $l_3$ are the same. On the contrary, the flatness of the probability distribution (2) and consequently the value of $EP^{(0)}_{\Gamma_j}$  depend, in the way described above, on the value of $k_0^j$.

We now consider a different situation in which it is meaningful to compare the values of $EP^{(0)}_{\Gamma_j}$ for different rooted graphs. We take as example the set of graphs $B = \{ \Gamma_{10}, \Gamma_{11}, \Gamma_{12},\Gamma_{13} \}.$  It is interesting to notice that the non-bipartite graphs $\Gamma_{11}, \Gamma_{12}$, and $\Gamma_{13}$ can be seen as generated from the bipartite graph $\Gamma_{10}$ by adding respectively one, two and three links, that is by increasing the connectivity of the subgraph $E'$. It is then interesting to see what happens to the maximum attainable value of the mode entanglement when the connectivity of the subgraph changes.
We refer to the figures (18) and (19), which describe the cases $\varepsilon >0$ and $\varepsilon <0$ respectively ($\ket {\psi_{in}}=\ket {1111}$). In figure (18) we can see that, for a given $\varepsilon >0$, the increase of the connectivity of the subgraph $E'$ has the effect of reducing the value of $EP^{(0)}_{\Gamma_j}$. On the contrary, when $\varepsilon <0$ the effect is opposite, see figure (19). What one could conclude by looking at equations (21)-(22) is that $EP^{(0)}_{\Gamma_j}$ should decrease when the connectivity of the subgraph increses, regardless of the sign of the self-interaction, in view of the dependance of $\rho_1^{(0)}$ and $\rho_2^{(0)}$ on $k_0/k$. But this contradicts the behaviour described in figure (19). In fact, it is the role of the sign of the self-interaction in $P$ that allows $EP^{(0)}_{\Gamma_j}$ to increase with the increase of the connectivity of the sub graph $E'$. We can see here the role of the interplay between the self-interaction and the tunnelling processes. 

We can now end our analysis by considering the case $N=L$ and $\ket{\psi_{in}}=\ket{N}\otimes_{j=0}^{L-1} \ket{0}_j.$ As a first step we can write the analitical expression of $P$ in terms of the graph properties as follows:
\be
P = \frac{4 \tau^2 k_0 N }{4 \tau^2 k_0 N  + [2(N-1)\varepsilon -2\tau \frac{l_3^0}{k_0}]^2},
\ee
where $k_0$ is the number of links starting from the root vertex, while $l_3^0$ is the number of loops of length three that include the root vertex.
Since for all the states in $\ket{I}$ $n_0=N-1$, for the determination of $EP^{(0)}_{\Gamma_j}$ in the perturbative regime the only terms of the probability distribution (2) that are relevant are $\rho_N^{(0)}$ and $\rho_{N-1}^{(0)}$. These can be simply written as:
\bae
\rho_N^{(0)} &=& 1 - P \\
\rho_{N-1}^{(0)} &=& P.
\eae
With the help of (24) we can now see that the result (23) hold also for the state $\ket{\psi_{in}}=\ket{N}\otimes_{j=0}^{L-1} \ket{0}_j.$ In fact, here it is  dependence of $P$ on $k_0$ that determines the behaviour described by (23).
Furthermor, also in this case it is possible to say that the $\Gamma_j$'s for which $EP^{(0)}_{\Gamma_j}$ is higher are those for which $k_0^j$ is maximum.

Finally we consider again the case $B = \{ \Gamma_{10}, \Gamma_{11}, \Gamma_{12},\Gamma_{13} \}$, i.e. the rooted  graphs that can be can be seen as generated from the bipartite $\Gamma_{10}$ by increasing the connectivity of the subgraph $E'$.  In this case the increase of the connectivity of the subgraph plays a role only if it implies the presence in the graph of loops of length three that include the root vertex. For the set $B$ we have that $l_3^0=0,1,2,3$ for $\Gamma_{10}, \Gamma_{11}, \Gamma_{12}$, and $\Gamma_{13}$ respectively. The interplay between the self-interaction and the tunnelling processes plays a role which is opposite with respect to the case
$\ket{\psi_{in}}=\ket{1111}$. In fact, in the case $\ket{\psi_{in}}=\ket{4000}$ we have that for a given $\varepsilon >0$, the increase of the connectivity of the subgraph $E'$ has the effect of increasing the value of $EP^{(0)}_{\Gamma_j}$. On the contrary, when $\varepsilon <0$ the effect is opposite.

\section{conclusions}
We have analyzed by means of numerical simulations 
the  mode-entanglement properties of of bosonic particles hopping over graph structures.
Starting from an initial unentangled
state bi-partite mode-entanglement of a graph vertex with respect the rest of the graph is  generated,
by allowing  for a finite time a tunneling along the graph edges.
We characterized the entangling power of a given hopping hamiltonian by the
maximum achieved during the dynamical evolution
by this entanglement.
We studied the dependence of this entangling power   by the local self-interaction parameters 
for all the graphs up to four vertices and for two different natural choices of the initial state.
Our results show the important role of that graph topology has in optimization of  entanglement production

The main steps and results,for both cases $N=L=3,4$ , can be summarized as follows.

\begin{itemize}

\item{}
We have classified the way in which a mode (bosonic site) can be linked to the "rest of the world" (other bosonic sites) in
terms of rooted graphs which are inequivalent from the poin of view of the mode entanglement of the reference mode (root).

\item{} When  the self-interaction term is absent ($=0$) the value of $EP_{\Gamma_j}^{(0)}$ does not depend on the graph
choosen 
but only on the initial state.

\item{}We have established for which of the configurations ($\Gamma_j, \ket{\psi_{in}}$) the free evolution of the system 
can lead to the optimal value of $EP_{\Gamma_j}^{(0)}$.

\item{}
We have given a general (analytical) justification of the different behaviour of 
$EP_{\Gamma_j}^{(0)}$ with respect to the change of sign in the selfinteraction parameter.
 This has been done by first observing that the graphs $\Gamma_j$ can be classified in {\it bi-partite} and 
{\it non bi-partite}. We have thus given an analitical justification of the fact that 
\[
EP^{(0)}_{\Gamma_j}(\varepsilon)=EP^{(0)}_{\Gamma_j}(-\varepsilon)\,\   \forall 
\varepsilon \in [0,20] 
\]
for all the {\it bi-partite} graphs. 
This relation, as it is highlited by our simulations, is not true for the {\it non bi-partite} ones.

\item{}
We have given a general explanation for the reduction of $EP_{\Gamma_j}^{(0)}$ when the self-interaction 
term becomes sufficiently greater than the hopping term. 
This has been done by adopting a two-levels system perturbative approach.

\item{}For the {\it non bi-partite} graphs, in the perturbative two-levels picture, we have given a 
detailed description of the following interesting features highlighted by our simulations. 
In the perturbative regime, the entangling power of the pair $(U_{\Gamma_j}, \ket{\psi_{in}})$ is: 
enhanced (decreased) by a positive (negative) self-interaction when, in the initial state, the particles are all localized in the reference mode ($n_0 = N$); the opposite holds true when the initial state is uniformly occupied ($n_i=1,\ \forall \ i=1,..L$).

\item{}
In the context of  the perturbative two-levels picture we have derived  analytical formulas that allow to link  $EP_{\Gamma_j}^{(0)}$ to the topological properties of the rooted graphs; this has been done in the general case $N=L$ and for both the initial states $\ket{\psi_{in}}=\ket{N}\otimes_{j=0}^{L-1} \ket{0}_j$ and $\ket{\psi_{in}}=\otimes_{j=0}^{L-1} \ket{1}_j$.
Starting from the simulations and with the help of this formulas we have compared and described the behaviour of $EP_{\Gamma_j}^{(0)}$ for different graphs.

\end{itemize}

We believe that a  natural --though not unique -- experimental framework in which the theoretical analysis 
carried over in this paper can find application is provided by systems of cold bosonic atoms in optical lattices.

\section{acknowledgments.} 
We are grateful to Radu Ionicioiu for useful comments.
Special thanks are due to Vittorio Penna for introducing us to the problem of
bosonic lattices and useful discussions. We gratefully acknowledge financial support by 
Cambridge-MIT Institute Limited and by the European Union project  TOPQIP

%\end{document}

\vspace{2cm}

\end{document}